\documentclass{conm-p-l}
\input{amssym.def} \input{amssym.tex} 
\renewcommand{\ll}{\label}
\newcommand{\be}{\begin{equation}}
\newcommand{\ee}{\end{equation}}
\newcommand{\bea}{\begin{eqnarray}}
\newcommand{\eea}{\end{eqnarray}}
\newcommand{\nn}{\nonumber}
\newcommand{\bib}{\bibitem}
\newcommand{\ci}{\cite}
\newcommand{\qm}{quantum mechanics}

\newcommand{\ca}{$C^*$-algebra}
\newcommand{\rep}{representation}
\newcommand{\irrep}{irreducible representation}
\newcommand{\Hs}{Hilbert space}
\newcommand{\sta}{$\mbox{}^*$-algebra}

\newcommand{\ovl}{\overline}
\newcommand{\wt}{\widetilde}
\newcommand{\til}{\tilde}
\newcommand{\raw}{\rightarrow}

\newcommand{\ot}{\otimes}

\newcommand{\la}{\langle}
\newcommand{\ra}{\rangle}

\newcommand{\x}{\times}

\newcommand{\BH}{{\frak B}({\mathcal H})}

\newcommand{\cin}{C^{\infty}}
\newcommand{\cci}{C^{\infty}_c}
\newcommand{\half}{\mbox{\footnotesize $\frac{1}{2}$}}

\newcommand{\Ah}{A_{\hbar}}
\newcommand{\lho}{\lim_{\hbar\rightarrow 0}}
\newcommand{\tsr}{T^*\Bbb R^n}

\newcommand{\inv}{^{-1}}
\newcommand{\Exp}{{\rm Exp}}

\newcommand{\q}{{\mathcal Q}_{\hbar}}
\newcommand{\CPW}{C^{\infty}_{\mbox{\tiny PW}}}
\newcommand{\al}{\alpha}

\newcommand{\gm}{\gamma}

\newcommand{\et}{\eta}

\newcommand{\io}{\iota}
\newcommand{\kp}{\kappa}
\newcommand{\lm}{\lambda}

\newcommand{\rh}{\rho}
\newcommand{\sg}{\sigma}

\newcommand{\ta}{\tau}

\newcommand{\phv}{\varphi}
\newcommand{\ch}{\chi}
\newcommand{\ps}{\psi}
\newcommand{\Ps}{\Psi}
\newcommand{\sP}{{\sf P}}

\newcommand{\Om}{\Omega}
\newcommand{\A}{ A}
\newcommand{\B}{{\frak B}}
\newcommand{\GC}{C}

\newcommand{\g}{{\frak g}}

\newcommand{\CA}{{\mathcal A}}

\newcommand{\F}{{\mathcal F}}

\renewcommand{\H}{{\mathcal H}}

\newcommand{\CO}{{\mathcal O}}

\newcommand{\CQ}{{\mathcal Q}}
\newcommand{\qw}{{\mathcal Q}_{\hbar}^W}

\makeatletter
\newskip\tempskip
\def\endproof{{\parfillskip24\p@ plus\@ne fil\@@par}\tempskip\prevdepth
  \ifdim\lastskip=\z@\tempskip\z@\else\vskip-\lastskip
    \ifdim\tempskip>4\p@ \tempskip.5\tempskip \else \tempskip\z@\fi\fi
  \nobreak\vskip-\baselineskip\vskip-\tempskip\noindent\hbox 
to\hsize{\hfill
    $\blacksquare$}\par\vskip\tempskip\vskip\abovedisplayskip\@doendpe}
\makeatother
\makeatletter
\newskip\tempskip
\def\endiproof{{\parfillskip24\p@ plus\@ne fil\@@par}\tempskip\prevdepth
  \ifdim\lastskip=\z@\tempskip\z@\else\vskip-\lastskip
    \ifdim\tempskip>4\p@ \tempskip.5\tempskip \else \tempskip\z@\fi\fi
  \nobreak\vskip-\baselineskip\vskip-\tempskip\noindent\hbox 
to\hsize{\hfill
    $\Box$}\par\vskip\tempskip\vskip\abovedisplayskip\@doendpe}
\makeatother

 \newcommand{\Ao}{A^0}
 \newcommand{\Pri}{\mbox{\rm Prim}} 
\newtheorem{Theorem}{Theorem}[section]
\newtheorem{Lemma}[Theorem]{Lemma}
\newtheorem{Proposition}[Theorem]{Proposition}
\newtheorem{Corollary}[Theorem]{Corollary}
\newtheorem{Definition}[Theorem]{Definition}
\newtheorem{Example}[Theorem]{Example}
\theoremstyle{remark}
\newtheorem{remark}[Theorem]{Remark}
\numberwithin{equation}{section}
\def\stackunder#1#2{\mathrel{\mathop{#2}\limits_{#1}}}
\def\d{\partial}
\newcommand\dint{\displaystyle \int }
\newcommand\dsum{\displaystyle \sum }
\newcommand\C{\mathbb C}
\newcommand\R{\mathbb R}
\newcommand\N{\mathbb N}
\newcommand\GG{\frak G}

\begin{document}
\title{Quantization of Poisson algebras associated to  Lie algebroids}
\author{N.P. Landsman}
\address{Korteweg-de Vries Institute for Mathematics, University of 
Amsterdam, Plantage Muidergracht 24, 1018 TV Amsterdam,The Netherlands}
\email{npl@wins.uva.nl}
\author{B. Ramazan}
\address{Institute of Mathematics of the Romanian Academy, Calea Grivi\c tei 
21, P.O. Box 1-764, 70700 Bucharest, Romania}
\email{ramazan@pompeiu.imar.ro}
\date{November 1, 1999}

\begin{abstract}
We prove the existence of a strict deformation quantization for the
canonical Poisson structure on the dual of an integrable Lie
algebroid. It follows that any Lie groupoid
\ca\ may be regarded as a result of a quantization procedure. 
The \ca\ of the tangent groupoid of a given Lie groupoid $G$ (with Lie
algebroid $A(G)$) is the
\ca\ of a continuous field of \ca s over $\R$ with fibers
$A_0=C^*(A(G))\simeq C_0(A^*(G))$ and $A_{\hbar}=C^*(G)$ for $\hbar\neq
0$. The same is true for the corresponding reduced \ca s. Our results
have applications to, e.g., transformation group \ca s, $K$-theory,
and index theory.
\end{abstract}

\maketitle
\tableofcontents
\section{Introduction}\ll{int}
The idea of ``quantization'' has evolved through a  number of
stages.  At the beginning of this century, one meant the fact that at
a microscopic scale certain physical quantities (like energy or
 angular momentum) assume only discrete values.  Such discreteness
is easily understood within the \Hs\ formalism of \qm, where
self-adjoint operators may or may not have a discrete spectrum, and is
no longer seen as the defining property of a quantum theory.

Since about 1925, the idea has referred to the passage from a classical to a
``corresponding'' quantum theory.  There are basically two ways to
describe this passage, starting either from the Lagrangian or from the
Hamiltonian version of classical mechanics. 

As first recognized by Feynman, the Lagrangian approach naturally
leads to the path-integral formulation of \qm, and writing down a path
integral may be seen as an act of quantization. All quantum-mechanical
observables are constructed from so-called transition amplitudes,
which correspond to the integration of certain functions with respect
to the path integral. Though intuitively appealing, it is often hard
to make this step  rigorous.  Through the nineties,
many ideas of Witten originated from  the use of
the path integral. 

Most mathematically precise work on quantization is based on the
Hamiltonian formulation of classical mechanics. It was initially
believed that the underlying mathematics consisted of symplectic
geometry, but since about 1976 it has been understood that classical
mechanics should be based on the concept of a Poisson algebra
\ci{Kir,Lic}. This is a commutative algebra with a Lie bracket that
turns each element of the algebra into a derivation with respect to
the commutative structure; see Definition \ref{defpoisal} below. Most
Poisson algebras one encounters in physics are of the form $\cin(P)$,
where $P$ is a so-called Poisson manifold; the commutative algebra
structure comes from pointwise multiplication, and the Lie bracket is
just the Poisson bracket $\{\, ,\,\}$.  Symplectic manifolds form a
special, nondegenerate case.

As first recognized by Heisenberg in 1925, the quantum-mechanical
observables of a given physical system should form a noncommutative
algebra; the noncommutativity leads to the uncertainty relations that
form the physical basis of \qm.  A general, heuristic theory
incorporating this idea was given in Dirac's book \ci{Dir}.  The
correct mathematical formalism of \qm, which still stands today, is
due to von Neumann
\ci{vN}, who created the abstract theory of \Hs s and self-adjoint 
operators for this purpose.  The only (but crucial) modification to
von Neumann's formalism has been to allow other \ca s than $\BH$ (the
algebra of all bounded operators on a \Hs\ $\H$) as algebras of
observables. This generalization corresponds to admitting
superselection rules; in the context of the quantization of
finite-dimensional systems this means that one incorporates Poisson
manifolds that are not symplectic in the underlying classical
theory. The examples in this paper are of such a form.

In the Hamiltonian formalism, quantizing a Poisson algebra $A^0$ of
classical observables amounts to finding a ``corresponding''
noncommutative associative algebra $A$ of quantum observables, as well
as a quantization map $\CQ:A^0\raw A$, subject to certain conditions.
Initially, practically all of quantization theory was based on a
single idea of Dirac \ci{Dir}, which he conceived in 1926 during a
Sunday walk near Cambridge. Namely, in \qm\ the role of the Poisson
bracket of the classical theory should be played by $1/i\hbar$ times
the commutator.  Here $\hbar$ is a small positive number, which in
physics is a constant of nature.  Hence if a classical observable $f$
is quantized by a quantum observable $\CQ(f)$, one expects that
``Dirac's condition'' $$(i\hbar)\inv[\CQ(f),\CQ(g)]=\CQ(\{f,g\})$$
holds at some fixed value of $\hbar$. In other words, the quantization
map $\CQ$ should be a Lie algebra homomorphism with respect to the
Poisson bracket and the (rescaled) commutator.

Geometric quantization \ci{Woo} was an attempt to make this idea
precise. Given a ``prequantizable'' symplectic manifold $S$ with
associated Poisson algebra $\cin(S)$, this approach produces a \Hs\
$\H(S)$ (the space of $L^2$-sections of the prequantization line
bundle $L(S)$ over $S$) and a Lie algebra homomorphism
$\CQ^{\mbox{\tiny pre}}$ from $\cin(S)$ to a certain algebra $\CA(S)$ of
unbounded operators on $\H(S)$ that are densely defined on the domain
of smooth sections of $L(S)$.

 The case where $S$ is a coadjoint orbit in the dual $\g^*$ of the Lie
 algebra $\g$ of a Lie group $G$ has been studied in particular
 detail, for the following reason.  Each element of $X\in \g$ defines a
 function $\til{X}$ on $\g^*$, and hence on $S$, by
 $\til{X}(\theta)=\theta(X)$. One has
\begin{equation}
 \label{lps} \{\til{X},\til{Y}\}=\widetilde{[X,Y]}, 
\end{equation}
so that $\CQ^{\mbox{\tiny pre}}$
 restricts to a Lie algebra homomorphism from $\g$ to $\CA(S)$.
 Assuming that this (infinitesimal) \rep\ of $\g$ is integrable to a
 unitary \rep\ $U$ of $G$, one finds that $U$ tends to be
 reducible. In order to obtain an irreducible sub\rep\ of $U$, one
 must restrict $\H(S)$. In the simplest cases one may simply project
 onto an irreducible subspace with orthogonal projection $P$ in the
 commutant of $U(G)$.  For general $f\in\cin(S)$, one then has to
 define $\CQ(f)=P\CQ^{\mbox{\tiny pre}}(f)P$; it is clear that this
 modification destroys the Lie algebra homomorphism property (except
 on the linear span of the $\til{X}$, since $[\CQ^{\tiny
 pre}(\til{X}),P]=0$). 

 A vast number of other methods has been proposed to achieve
 irreducibility; see \ci{Kir} for a recent overview. The conclusion is
 that one cannot achieve irreducibility of $U(G)$ while preserving the
 Lie algebra homomorphism property on all smooth functions. In other
 examples than coadjoint orbits, one finds that the map
 $\CQ^{\mbox{\tiny pre}}$ is unsatisfactory for other (though related)
 reasons, the conclusion being the same: a satisfactory quantization
 map is only a Lie algebra homomorphism in the stated sense on some
 subspace of the Poisson algebra of classical observables.

The way out of this dilemma emerged in the seventies, mainly as a
consequence of the work of Berezin \ci{Ber}, Vey \ci{Vey}, and  Bayen et al.\
\ci{BFFLS}. It is, quite simply, to require Dirac's condition only
asymptotically, that is, for $\hbar\raw 0$.  This, of course,
necessitates the dramatic step of defining the quantization data $A$
and $\CQ$ for a family $I\subseteq\R$ of values of $\hbar$ that
contains 0 as an accumulation point. Hence, given a Poisson algebra
$A^0$, one now needs to find a family of algebras $\{\Ah\}_{\hbar\in
I}$ and maps $\q:A^0\raw\Ah$, with $\CQ_0={\rm id}$, such that
$(i\hbar)\inv [\q(f),\q(g)]\raw\q(\{f,g\})$ in some sense. A method
that accomplishes this is generically referred to as a deformation
quantization.

There are (at least) two ways to make rigorous sense of the above
convergence.  The oldest, introduced in the context of quantization
theory in \ci{BFFLS}, is to define $\Ah$ and $\q$ as formal power
series in $\hbar$.  This method, called formal deformation
quantization, remains the most popular; see \ci{Kon} for a recent
highlight.

 The second approach, which is based on choosing  the $\Ah$ to be
 \ca s, was introduced by Rieffel \ci{Ri2}-\ci{Ri7}.  Here $\hbar$
 is simply a real number rather than a formal deformation parameter,
 and one imposes Dirac's condition asymptotically in norm; see
 Definition \ref{defqua} below.  It is then mathematically natural
 and physically meaningful to require that the $\Ah$ form a continuous
 field of \ca s over the index set $I\ni\hbar$. In section \ref{Sdq} we 
 give a precise formulation of this approach to deformation
 quantization, introducing two appropriate generalizations of Rieffel's
original definition that lie at the basis of the results in the present paper.

Having discussed the development of modern quantization theory in
broad outline, we now turn to the class of classical systems (that is,
Poisson manifolds) that are quantized in this paper. Our motivation
comes from a number of directions. 

The first is the quantization theory of coadjoint orbits discussed
above. Although the attempt to quantize individual orbits is only
successful in special cases, one could try to quantize all coadjoint
orbits of a given Lie group at one stroke by quantizing $\g^*$ (or
$\cin(\g^*)$) as a whole. To do so, one starts from the fact that
$\g^*$ is canonically a Poisson manifold when equipped with the
so-called Lie--Poisson structure that may be defined by
(\ref{lps}). As first shown by Rieffel \ci{Ri4} in the special case
that $G$ is exponential and nilpotent, one may construct a
quantization of $\cin(\g^*)$ in which $\Ah=C^*(G)$ for all $\hbar\neq
0$, and the maps $\q$ are essentially given by the pullback of the
exponential map from $\g$ to $G$, rescaled by $1/\hbar$.  We show
that this works without any assumption on the Lie group $G$.  The idea
that coadjoint orbits in $\g^*$ are quantized by unitary
\irrep s of $G$ then re-enters through the back door, as follows. The
well-known correspondence between nondegenerate
\rep s of $C^*(G)$ and unitary  \rep s of $G$, preserving irreducibility, 
is matched by the fact that the \irrep s of $\cin(\g^*)$ as a Poisson
algebra \ci{La3} are precisely given by the coadjoint orbits of $G$
(or covering spaces thereof). In general, there is no correspondence
through quantization between the \irrep s of $\cin(\g^*)$ and the
\irrep s of $C^*(G)$; the correspondence through (deformation)
quantization is between the algebras in question themselves.

Secondly, one may quantize a particle moving on a Riemannian manifold
$M$ (so that the pertinent Poisson manifold is the cotangent bundle
$T^*M$, which is symplectic) by taking $\Ah$ to be the \ca\
$\B_0(L^2(M))$ of compact operators on $L^2(M)$ for all $\hbar\neq 0$,
and defining a quantization map $\q$ in terms of the geodesics on $M$.
This result may be generalized by coupling the particle to an external
Yang--Mills field with gauge group $H$; one then starts from a
principal $H$-bundle $\sf P$, so that $M={\sf P}/H$.  The Poisson
manifold describing the classical theory is then given by
$P=(T^*\sP)/H$, which is not symplectic unless $H$ is discrete, and
when $H$ is compact the quantum algebra of observables turns out to be
$\B_0(L^2(\sP))^H$; see \ci{La1,La3} for details and the
generalization to noncompact $H$.  Taking $\sP=H=G$ actually
reproduces the previous example, since $(T^*G)/G\simeq \g^*$ as
Poisson manifolds, whereas for compact Lie groups one has
$\B_0(L^2(G))^G\simeq C^*(G)$ (and analogously for the noncompact
case).  The quantization maps in question are also compatible with
this specialization.

Thirdly, in the early eighties Lie groupoids and locally compact
groupoids started to play a role in \ca s as a result of the work of
Connes and of Renault, whereas in the late eighties Lie groupoids
entered symplectic and Poisson geometry, as well as quantization
theory, through the idea of a symplectic groupoid (cf.\
\ci{Wei98}). Against this background, it was an obvious idea that the
above quantizations ought to be interpretable in terms of Lie
groupoids.  The key to this possibility lies in the fact the passage
from a Lie groupoid to its convolution \ca\ has a classical
analogue. Namely, like a Lie group, a Lie groupoid $G$ has an
associated ``infinitesimal'' object, its Lie algebroid \ci{P0} $A(G)$,
which is a vector bundle over the base space $G_0$ of $G$. The dual
bundle $A^*(G)$ of $A(G)$ admits a canonical Poisson structure
\ci{coste,Cou}, so that eventually one may associate a Poisson algebra
$\cin(A^*(G))$ to $G$.

All threads now come together \ci{La1}, in that in each of the three
classes of examples above the classical Poisson algebra is of the form
$\cin(A^*(G))$ and the quantum \ca\ is of the form $C^*(G)$, where $G$
is some Lie groupoid. This is obvious in the first example. In the
second, one takes $G$ to be the pair groupoid $G=M\x M$ on $M$, whose
Lie algebroid is the tangent bundle $TM$; the Poisson structure on the
dual bundle $T^*M$, which is nothing but the cotangent bundle of $M$,
is just its usual canonical structure (which is symplectic). 
See also \cite{CCFGRV}.  Finally,
in the third example $G$ is the so-called gauge groupoid of the
principal bundle $\sP$ \ci{Mac}, with associated Lie algebroid
$(T\sP)/H$.

Based on these examples, it was conjectured in \ci{La2} that one should obtain a deformation
quantization of this type for any Lie groupoid $G$. For formal deformation quantization
a version of this conjecture was proved in \ci{nistor}, and a proof in \ca ic deformation
quantization above appeared in \ci{La3,La4,Ra1}.

In this paper we give a precise and complete formulation of all
mathematical concepts surrounding this class of \ca ic deformation
quantizations, and prove two technically distinct versions of the
above conjecture, each of which has its merits.  The generalization of
Connes's tangent groupoid \ci{connes} $\hat{G}$ to arbitrary Lie
groupoids $G$ \ci{hilsum,Wei89} plays an important role.  A
significant intermediate result is that $C^*(\hat{G})$ is the \ca\ of
a continuous field of \ca s over $\R$ with fibers $A_0=C^*(A(G))\simeq
C_0(A^*(G))$ and $A_{\hbar}=C^*(G)$ for $\hbar\neq 0$. The same is true
for the corresponding reduced \ca s. See section \ref{tg}.

The sections of the paper are listed in the table of contents
following the abstract.  Our main results, Theorems \ref{main1} and
\ref{main2}, are stated at the end of section \ref{Sdq}.  The
remainder of the paper develops the proofs of these theorems.
\section{Lie groupoids and Lie algebroids}\label{LgLa}
This section is a brief review of our objects of study, mainly
intended to establish our notation.  We assume that the reader is
familiar with the basic theory of locally compact groupoids (cf.\
\ci{Re1}), which we always assume to be second countable.  We denote a
groupoid as a whole by $G$, the base is called $G_0$, which is seen as
a subspace of the arrow space (or total space) $G_1$ under the
inclusion map $\io$. The source and range projections are called $s$
and $r$, respectively, and for $x\in G_0$ we put
\begin{equation}
G^x=r\inv(x)\subset G_1.
\end{equation}
The space of composable arrows is $G_2=\{(\gm,\gm')\in G_1\x G_1\mid
r(\gm')=s(\gm)\}$.
\begin{Definition}
A Lie groupoid is a locally compact groupoid $G$ for which $G_1$ and
$G_0$ are manifolds, $s$ and $r$ are surjective submersions, and
multiplication and inclusion are smooth maps.
\end{Definition}

It follows that $\io$ is an immersion, that the inverse is a
diffeomorphism, that $G_2$ is a closed submanifold of $G_1\x G_1$,
that the fibers $r\inv (x)$ and $s\inv(x)$ are submanifolds of $G_1$
for all $x\in G_0$, and that all isotropy groups are Lie groups.  The
basic reference on Lie groupoids is Mackenzie's book \ci{Mac}; also
see \ci{connes,La3}.

A Lie group defines its Lie algebra; one may also study Lie algebras
in their own right. The situation for Lie groupoids is similar.
\begin{Definition} \label{defloid}
A  Lie algebroid  on a manifold
$M$ is a vector bundle $E$ over $M$, which apart from the
bundle projection $\ta:E\raw M$ is equipped with a vector bundle map
$\rh:E\raw TM$ (called the anchor), as well as
with a Lie bracket $[\, ,\,]_{E}$ on the space $\cin(M,E)$ of smooth
 sections of $E$, satisfying \begin{equation}  \label{anchor1}\rh\circ
[X,Y]_{E}=[\rh\circ X,\rh\circ Y], \end{equation}
where the right-hand side is the usual commutator of vector fields on
$\cin(M,TM)$, and 
\begin{equation} \label{anchor2} [X,f Y]_{E}=f [X,Y]_{E}+ ((\rh\circ
X) f) Y  
\end{equation} for all $X,Y\in\cin(M,E)$ and
$f\in\cin(M)$.  We generally omit the suffix $E$ on the Lie bracket.
\end{Definition}

It is part of the definition of a bundle map that the anchor is
fiber-preserving and linear on each fiber.  This concept is due to
Pradines \ci{P1}. The basic reference on Lie algebroids is \ci{Mac};
also see \ci{La3}.  The simplest examples of Lie algebroids are
tangent bundles $TM$, where $\rh$ is the identity, and Lie algebras,
for which $M$ is a point.

We now  explain how one may associate a Lie algebroid $A(G)$ with a
given Lie groupoid $G$ \ci{P0,Mac,La3}.
A left-invariant vector field
$\xi^L$ on $G_1$ is a vector field satisfying $Tr(\xi^L)=0$ and
$TL_{\gm}(\xi^L)(\gm')=\xi^L(\gm\gm')$ for all $(\gm,\gm')\in G_2$.
Here $L_{\gm}:G^{s(\gm )}\rightarrow G^{r(\gm )}$
is defined for each $\gm\in G_1$  by 
\begin{equation}
L_{\gm}(\gm _1)=\gm \gm _1. \label{defLgm}
\end{equation}
Note that the second condition is well-defined because of the first
one.  The space of all smooth left-invariant vector fields on $G_1$ is
denoted by $\cin(G_1, TG_1)^L$, which is easily shown to be a Lie
algebra under the usual commutator borrowed from $\cin(G_1,TG_1)$.
Also, a left-invariant vector field is obviously determined by its
values on the unit space $G_0$.  The tangent bundle of $G_1$ at the
unit space has a decomposition
\begin{equation}
\label{defTr}
T_xG_1  =  T_xG_0\oplus (\ker Tr)_x,  
\end{equation}
where $\ker Tr\subset TG_1$ is the vector bundle over $G_1$ whose
fiber $(\ker Tr)_{\gm}$ above $\gm\in G_1$ is the kernel of the
derivative $T_{\gm}r:T_{\gm}G_1\raw T_{r(\gm)}G_0$ of the range
projection $r:G_1\raw G_0$. The special case $\gm=x\in G_0$ occurs in
(\ref{defTr}).
\begin{Definition}\label{deflbroid}
The Lie algebroid $A(G)$ of a Lie groupoid $G$ is given by
the following:
\begin{enumerate}
\item
The vector bundle $A(G)$ over $G_0$ is the normal bundle 
defined by the embedding $G_0\hookrightarrow G_1$; accordingly, the bundle
projection $\ta:A(G)\raw G_0$ is given by $s$ or $r$ (which coincide on $G_0$).
\item
Identifying the normal bundle with $(\ker Tr)|_{G_0}$ by
(\ref{defTr}), the anchor is given by $\rh=Ts:\ker Tr\raw TG_0$.
\item Identifying a section of the normal bundle with
 an element of $\cin(G_1, TG_1)^L$ through the previous item, the Lie
 bracket $[\, ,\,]_{A(G)}$ is given by the commutator.
\end{enumerate}\end{Definition}

The required equality (\ref{anchor1}) is automatically satisfied (as
it holds for all vector fields on $G_1$). We leave the verification of
(\ref{anchor2}) to the reader.

It follows from this definition that
a Lie algebra $\g$ is the Lie algebroid of a Lie group $G$, and that
the tangent bundle $TG_0$ is the Lie algebroid of the pair groupoid
$G_1=G_0\x G_0$.

For Lie groups one has an exponential map from the Lie algebra to the
group.  For manifolds $M$ with connection the exponential map is
defined on the tangent bundle $TM$, which it maps into $M$. As
indicated by Pradines \ci{P2}, these are special cases of a
construction that holds for general Lie groupoids $G$, provided its
Lie algebroid $A(G)$ is endowed with a connection. Following
\ci{La3,La4}, we here present a slightly different construction that
is more suitable for our application to deformation quantization; also
see \ci{nistor}.

First note that the
vector bundles $\ker Tr$ and $s^*A(G)$ (over $G_1$) are isomorphic via the
map $TL_{\gm}$, applied fiberwise.
Hence the connection on $A(G)$, with associated horizontal lift
 $\ell^{A(G)}$, yields  a connection
 on $\ker Tr$ through pull-back.  Going through the
 definitions, one finds that the associated horizontal lift $\ell$ of
 a tangent vector $X=\dot{\gm}=d\gm(t)/dt_{t=0}$ in $T_{\gm}G_1$ to
 $Y\in T_{\gm}^rG_1$ is \begin{equation} \ell_Y(\dot{\gm})=\frac{d}{dt}
 [L_{\gm(t)*}
 \ell^{A(G)}_{TL_{\gm\inv}(Y)}(s(\gm(t)))]_{t=0},\ll{assconncom} \end{equation}
 which is an element of $T_Y(\ker Tr)$ (here $\ell^{A(G)}(\ldots)$ lifts
 a curve).

Since the bundle $\ker Tr\raw G_1$ has a connection, one can define
parallel transport $X\raw X(t)$ on $\ker Tr$ in precisely the same way as
on a tangent bundle with affine connection. That is, the flow $X(t)$
is the solution of \begin{equation} \dot{X}(t)=\ell_{X(t)}(X(t))
\ll{dotXtis} \end{equation} with initial condition $X(0)=X$. The
projection of $X(t)$ to $G_1$ is a ``geodesic'' $\gm_X(t)$.
\begin{Definition}\ll{defExpL}
The left exponential map $\Exp^L:A(G)\raw G_1$ is defined by \begin{equation}
\Exp^L(X)=\gm_{X}(1)=\til{\ta}(X(1)), \ll{ExpL}\end{equation}
where $\til{\ta}:\ker Tr\raw G$ is the restriction of the bundle projection $TG_1\raw G_1$
to $\ker Tr$.
 It is assumed that
 the geodesic flow $X(t)$ on $\ker Tr$ (defined by the
connection on $\ker Tr$ pulled back from the one on $A(G)$) is defined
at $t=1$. If not, $\Exp^L(X)$ is undefined.
\end{Definition}

There is a symmetrized version of $\Exp^L$ that plays a key role in
our quantization theory.  First note that for all $X\in A(G)$ for which
$\Exp^L(X)$ is defined one has \begin{equation}
r(\Exp^L(X))=\ta(X); \ll{tatExpL} \end{equation} recall that $\ta$ is
the bundle projection of the Lie algebroid. To derive this, note that
$r(\gm_X(0))=\ta(X)$ by construction, and $dr(\gm_X(t))/dt=Tr(X(t))=0$.

We combine this with the obvious $\ta(\half X)=\ta(-\half X)$ to infer
that
$$r(\Exp^L(\half X))=r(\Exp^L(-\half X))=s(\Exp^L(-\half
X)\inv).$$ 
Thus the (groupoid) multiplication in (\ref{ExpW}) below is
well-defined.
\begin{Definition}\ll{defExpW}
The Weyl exponential map $\Exp^W:A(G)\raw G_1$ is defined by \begin{equation}
\Exp^W(X)=\Exp^L(-\half X)\inv \Exp^L(\half X).\ll{ExpW} \end{equation}
\end{Definition}
 
The following well-known result is a form of the tubular neighbourhood
theorem.
\begin{Proposition}\ll{tntoid}
The maps $\Exp^L$ and $\Exp^W$ are diffeomorphisms from a
neighbourhood $V$ of $G_0\subset A(G)$ (as the zero section) to a
neighbourhood $W$ of $G_0$ in $G_1$, such that $\Exp^L(x)=\Exp^W(x)=x$
for all $x\in G_0$.
\end{Proposition}

Cf.\  \ci{La3,La4} for a proof.
\section{The $C^*$-algebra of a Lie groupoid}\label{CLg}
In this section we show that to every Lie groupoid $G$ one can
associate a $C^*$-algebra $C^*(G)$.  This $C^*$-algebra was introduced
by Bigonnet \cite{Bg}, but the idea to use half densities for an
intrinsic construction of the $C^*$-algebra of a Lie groupoid first
appeared in Connes's work \cite{Cn1} for the special case of the
holonomy groupoid of a foliation. Connes also proposed a general
definition in \ci{connes}; this only works if fiber isomorphisms of
the type discussed below are understood.  The construction we give in
this paper is due to Renault \cite{Re4}; it is essentially the same as
that in \cite{La2,La3}. Also see \ci{CW}.  Further details on
densities as used below may be found in \cite{treves,La3}.

Let $V$ be an $n$-dimensional vector space and let $\alpha$ be a fixed
real number.  A density of weight $\alpha$ on $V$ is a function $\rho
: \wedge ^n(V)\rightarrow {\C}$ that satisfies $$\rho (\lambda T)=\mid
\lambda \mid ^{\alpha} \rho (T)$$ for every $\lambda \in {\C}$ and $T\in 
 \wedge ^n(V)$.  The $1$-dimensional vector space of the densities of
 weight $\alpha$ on $V$ is denoted by $|\Om | ^{\alpha}(V)$. If a
 density is positive at a point, then it is positive everywhere on
 $\wedge ^n(V)\backslash \lbrace 0
\rbrace$. Hence one may unambiguously speak of positive densities.
 We have the canonical isomorphisms 
$|\Om |^{\alpha +\beta}(V)\simeq |\Om |^{\alpha}(V)\otimes 
|\Om |^{\beta}(V)$ and $|\Om |^{\alpha}(V\oplus W)\simeq |\Om 
|^{\alpha}(V)\otimes |\Om |^{\alpha}(W)$, where  $W$ is another vector space.

Denote the parallelepiped generated by a family $v_1,\ldots,v_n$ of
vectors in $V$ by $\left[ v_1,\ldots,v_n\right]$.  Given a positive
$1$-density $\rho$ on $V$, the translation invariant measure $\mu$ on
$V$ that satisfies $\mu (\left[ v_1,\ldots,v_n\right])=\rho (v_1\wedge
\ldots\wedge v_n)$ for every family $\left\{ v_1,\ldots,v_n\right\}$
of vectors is said to be the measure generated by $\rh$.  If $\left\{
e_1,\ldots,e_n\right\}$ is a basis of $V$ and $f\in L^1(V,\mu )$, then
\begin{equation}\label{calculintegral}
\dint f\left(\stackrel{n}{\stackunder{i=1}{\dsum }}x_ie_i\right)\rho 
(e_1\wedge \ldots\wedge e_n)dx=\dint fd\mu .
\end{equation}

We now generalize this construction to fibers in vector bundles.  Let
$E$ be a vector bundle on $M$, with fibers $E_x$ over $x\in M$. For
each $x\in M$, consider $|\Om |^{\alpha}(E_x)$, the $1$-dimensional
vector space of densities of weight $\alpha$ on the vector space
$E_x$. The vector bundle over $M$ with fibers $|\Om |^{\alpha}(E_x)$
is denoted by $|\Om |^{\alpha}(E)$. The sections of the vector bundle
$|\Om |^{\alpha}(E)$ over $M$ are called $\alpha$-densities. An
$\alpha$-density $\mu \in C^{\infty}(M,|\Om |^{\alpha}(E))$ is called
positive if $\mu (x)$ is a positive density for every $x\in M$. The
function associated to $\mu$ in the local frame $e=(e _1,\ldots,e _p)$
of $E$ on $U\subset M$ is the map $\mu _e\in C^{\infty}(U)$ defined by
$\mu _e (x)=\mu (x)\left( e_1(x) \wedge \ldots\wedge e_p (x)\right)$.

The set of $\alpha$-densities on $E$ is a free $1$-dimensional module over 
the ring of functions defined on $M$. Moreover, $C^{\infty}(M,|\Om |^{
\alpha}(E))$ is a free $1$-dimensional module over $C^{\infty}(M)$, and 
choosing a positive smooth density we can identify  densities with 
functions.

Now let $G$ be a Lie groupoid with  Lie algebroid $A(G)$; recall from Definition
\ref{deflbroid} that $A(G)$ is
a vector bundle over $G_0$. For every real number $\alpha$, one may
associate the vector bundle $|\Om |^{\alpha }(A(G))$ to $A(G)$.  In
addition, form the vector bundle $|\Om |^{\alpha}(
\ker Tr)$ of $\al$-densities associated to the vector 
bundle $\ker Tr\subset TG_1$.
(defined after (\ref{defTr})).

The groupoid $G$ acts (from the left)  on the bundle $|\Om |^{\alpha}(\mbox{
Ker}\,Tr)$. To define this action we need a further bit of notation.
Let $V,W$ be vector spaces with dim $V$=dim $W$. For every linear map 
$f:V\rightarrow W$, define $f^{\Om}:|\Om |^{\alpha}(F)\rightarrow |\Om 
|^{\alpha}(E)$ on $\rho \in \left| \Om \right| ^{\alpha}(F)$, 
$v_1,v_2,\ldots,v_n\in V$ by $(f^{\Om}\rho )(v_1\wedge \ldots\wedge v_n)=\rho 
(f(v_1)\wedge\ldots\wedge f(v_n))$. Obviously, if $\rho$ is a positive 
density then $f^{\Om}\rho$ is  positive, too.

To the derivative  $T_{\gm \gm_1}L_{\gm ^{-1}}:T_{\gm\gm _1}
G^{r(\gm )}\rightarrow T_{\gm _1}G^{s(\gm )}$ of $L_{\gm}$ 
(cf.\ (\ref{defLgm})) one associates the
corresponding  map between the spaces of densities 
$(T_{\gm \gm _1}L_{\gm ^{-1}})^{\Om}:\left| \Om \right| 
^{\alpha}(T_{\gm _1}G^{s(\gm )})\rightarrow |\Om |^{\alpha}(T_{
\gm \gm _1}G^{r(\gm )})$. For simplicity, we denote 
$(T_{\gm \gm _1}L_{\gm ^{-1}})^{\Om}$ by $\gm$, and have 
an action of $G$ on $|\Om |^{\alpha}(\ker Tr)$ given by  
$\gm :|\Om |^{\alpha}(T_{\gm _1}G^{s(\gm )})\rightarrow |\Om 
|^{\alpha}(T_{\gm \gm _1}G^{r(\gm )})$, where $\gm _1
\in G^{s(\gm )}$.

Denote the vector bundle $r^*|\Om |^{1/2}(A(G)) 
\otimes s^*|\Om |^{1/2}(A(G))$ over $G$ by  $\GG$. The fiber of $\GG$ over 
$\gm \in G$ is $\GG _{\gm}=|\Om |^{1/2}(T_{r(\gm )}
G^{r(\gm )})\otimes |\Om |^{1/2}(T_{s(\gm )}G^{s(\gm )})$. 
\begin{Definition}\label{id}
The convolution algebra of a Lie groupoid $G$ is the space
$C_c^{\infty }(G_1,\GG )$ of smooth compactly supported sections of
the vector bundle $\GG$ over $G_0$, with product $f*g$ of $f,g\in
C_c^{\infty }(G_1,\GG )$ defined by
\begin{equation}\label{Gconvolution}
(f*g)(\gm )=\stackunder {G^{r(\gm )}}{\dint }(\mbox{\rm id}\otimes \gm _1)
f(\gm _1)(\gm _1\otimes \mbox{\rm id})g(\gm _1^{-1}\gm ),
\end{equation}
where id is the identity map, and involution given by
\begin{equation}
f^*(\gm )=\widetilde{f(\gm ^{-1})}.
\end{equation}
 Here the map 
 $$ \sim:\left| \Om \right| ^{1/2}(T_xG^x)\otimes 
\left| \Om \right| ^{1/2}(T_yG^y)\rightarrow \left| \Om \right| 
^{1/2}(T_yG^y)\otimes \left| \Om \right| ^{1/2}(T_xG^x)$$ is defined by
$a(x)\otimes b(y)\mapsto \overline {b(y)\otimes a(x)}.$
\end{Definition}

It is not difficult to show that these expressions are well defined.
Let us show, for example, that the integral in (\ref{Gconvolution}) is
well defined. We extend the action of $G$ on the bundle $|\Om
|^{1/2}(\ker Tr)$ to the tensor product in an obvious way.  For
$f,g\in C_c^{\infty }(G_1,\GG )$ we see that $(\mbox{\rm id}\otimes
\gm _1)f(\gm _1)$ is an element of $|\Om |^{1/2}(T_{r(
\gm _1)}G^{r(\gm _1)})\otimes |\Om |^{1/2}(T_{\gm _1}G^{r(
\gm _1)})$, and that $(\gm _1\otimes \mbox{\rm id})g(\gm ^{-1}_1\gm )$ 
lies  in 
$|\Om |^{1/2}(T_{\gm _1}G^{r(\gm _1)})\otimes |\Om |^ {1/2}(T_{s(\gm
)}G^{s(\gm )})$.  The integrand in (\ref{Gconvolution}) is then an
element of $|\Om |^{1/2}(T_{r(\gm )}G^{r(\gm )})\otimes |\Om |^1(
T_{\gm _1}G^{r(\gm _1)})\otimes |\Om |^{1/2}(T_{s(\gm )}G^{s(
\gm )})$, so that it may be integrated as a 1-density on the manifold 
$G^{r(\gm )}$. Hence we obtain an element of $|\Om |^{1/2}(
T_{r(\gm )}G^{r(\gm )})\otimes |\Om |^{1/2}(T_{s(\gm )}G^{s(
\gm ) })$.
\begin{Proposition}\label{isodensit}
The map $\lambda \mapsto \lambda |_{G_0}$ is a bijection between
the left invariant sections of $|\Om |^{\alpha}(\ker Tr)$ and the 
sections of $|\Om |^{\alpha }(A(G))$. This bijection preserves 
smoothness as well as positivity.
\end{Proposition}

\begin{proof}
The restriction is well defined, since $|\Om | ^{\alpha }(\ker Tr)$
contains $|\Om |^{\alpha }(A(G))$. To prove that the map in question
is bijective, it is sufficient to check that its inverse is
$\rh\mapsto \lm$, where $\rho\in|\Om |^{\alpha }(A(G))$ and $\lambda
(\gm )=\gm \cdot \rho (s(\gm ))$.
\end{proof}

We now define the groupoid \ca\ $C^*(G)$ and its reduced counterpart
$C^*_r(G)$.  Firstly, to endow the convolution $\mbox{}^*$-algebra
$C^{\infty }_c(G_1,\GG )$ with a $C^*$-norm we consider its
$\mbox{}^*$-representations on a \Hs.  Let ${\mathcal R}$ be the set
of all $\mbox{}^*$-representations $\pi(C^{\infty }_c(G_1,\GG ))$ for
which $f\mapsto \langle \xi ,\pi(f)\eta \rangle $ is a Radon measure
on $C^{\infty }_c(G_1,\GG )$ for each pair $\xi ,\eta \in H$.  One
then defines a $C^*$-norm on $C^{\infty }_c(G_1,\GG )$ by
\begin{equation}
\|f\|=\stackunder{\pi\in {\mathcal R}}{
\mbox{sup}}\| \pi(f)\|;
\end{equation}
cf.\ \cite{Re1}. The completion
of $C^{\infty }_c(G_1,\GG )$ in this norm is the $C^*$-algebra $C^*(G)$. 

Secondly, for every $x\in G_0$ we define an involutive representation
$\pi_x$ of $C^{\infty }_c(G_1,\GG )$ on the Hilbert space $L^2(G^x)$
of half-densities on the manifold $G^x$ by \begin{equation}
\lbrack \pi _x(f)\xi \rbrack (\gm )=
\stackunder{G^x}{\dint}(\mbox{id}\otimes \gm _1)f(\gm _1)(\gm _1
\otimes \mbox{id})\xi (\gm _1^{-1}\gm ), \end{equation} for $\gm \in G^x$,
where
$\xi \in L^2(G^x)$.
Then $\|f\|_r=\stackunder{x\in G_0}{\mbox{sup}}\|\pi _x(f)\|$ is a norm 
(cf.\ \cite{Re1}), and the completion of $C^{\infty }_c(G_1,\GG )$ in
this norm is the reduced $C^*$-algebra $C_r^*(G)$.

For an amenable groupoid $G$ we have $C^*_r(G)=C^*(G)$.
For the proof of this result and a detailed discussion of amenable groupoids
see \cite{AR}.

Recall the following concept (Def.\ I.2.2 in \ci{Re1}).
\begin{Definition}
 A left Haar system on a locally compact groupoid $G$ is a family of
 measures $(\lambda^x)_{x\in G_0}$ such that the support of $\lm^x$ is
 $G^x$, the system is left-invariant under the map $L_{\gm}:
 G^{s(\gm)}\raw G^{r(\gm)}$, and for each $\phi\in C_c(G_1)$, the map
\begin{equation}
x \mapsto 
\lambda (\phi )(x)=\dint \phi \,d\lambda ^x \ll{lHs}
\end{equation}
 defines a continuous function on $G_0$. 

If $G$ is a Lie groupoid, we say that the Haar system is smooth if
each such function is smooth for $\phi\in C_c^{\infty }(G_1)$.
\end{Definition}

As explained in \ci{Re1},  one can associate 
\ca s $C^*(G,\lm)$ and $C_r^*(G,\lm)$ to a locally compact groupoid $G$ with
 Haar system $(\lm^x)$.  Recall that convolution and involution are
 given on the dense subalgebra $C_c(G_1)$ by
\bea
f* g(\gm) & = & \int_{G^{s(\gm)}}
 f(\gm \gm_1) g(\gm_1\inv)d\lm^{s(\gm)}(\gm_1) ; 
\ll{oldconv} \\ 
f^*(\gm) & = & \ovl{f(\gm\inv)}.\ll{oldinv} \eea This turns $C_c(G_1)$
into a $\mbox{}^*$-algebra, which we denote by $C_c(G,\lm)$.
Similarly, if $G$ is smooth one has the \sta\ $\cci(G,\lm)$.  The \ca
s $C^*(G,\lm)$ and $C_r^*(G,\lm)$ are completions of $C_c(G,\lm)$ or
$\cci(G,\lm)$ in suitable $C^*$-norms.

Not every locally compact groupoid admits a left Haar system; in the theory of
groupoid \ca s one therefore usually postulates its existence. 
 For Lie groupoids the situation is more favourable.
\begin{Proposition}\label{identification}
Any Lie groupoid $G$ admits a smooth left Haar system $(\lm^x)_{x\in
G_0}$.  The associated convolution $\mbox{}^*$-algebra $\cci(G,\lm)$
is isomorphic to $\cci(G_1,\GG)$, and the associated \ca s
$C^*(G,\lm)$ and $C_r^*(G,\lm)$ are isomorphic to $C^*(G)$ and
$C^*_r(G)$, respectively.
\end{Proposition}
\begin{proof}
Let $\lambda$ be a smooth positive section of $|\Om |^1(A(G))$. By
Proposition \ref{isodensit}, $\lambda$ can be extended to a smooth,
positive, and left invariant section $\lambda _G$ of $|\Om |^1(\ker
Tr)$. Then $\lambda ^{1/2}_G$ is a smooth left invariant section of
$|\Om |^{1/2} (\ker Tr)$, and $\gm \mapsto \lambda ^{1/2}_G(r(\gm
))\otimes \lambda ^{1/2}_G (s(\gm ))$ is a smooth section of
$\GG$. For every section $f\in C_c^{\infty }(G_1,\GG )$ there exists a
function $f_{\lambda }\in C_c^{
\infty }(G_1)$ such that
\begin{equation}\label{identif}
f(\gm )=f_{\lambda }(\gm )\lambda ^{1/2}_G
(r(\gm ))\otimes \lambda ^{1/2}_G(s(\gm )).
\end{equation}
An easy calculation shows that for $f,g\in C_c^{\infty }(G_1,
\GG )$ we have
$$(f*g)(\gm)=(f_{\lambda }*g_{\lambda })(\gm )\lambda ^{1/2}_G(r(
\gm ))\otimes \lambda ^{1/2}_G(s(\gm )),$$ where 
\begin{equation}\label{lambda_convolution}
(f_{\lambda }*g_{\lambda })(\gm )=\stackunder{G^{r(\gm)}}{\dint }f_{
\lambda }(\gm _1)g_{\lambda }(\gm _1^{-1}\gm )\lambda _G(\gm _1).
\end{equation}
The right-hand side of the last equality is well defined as the
integral of the $1$-density $f_{\lambda }(\gm _1)g_{\lambda }(\gm
_1^{-1}\gm )\lambda _G(\gm _1)$ over the manifold $G^{r(\gm )}$.  It
is also easy to see that
\begin{equation}\label{star}
f^*(\gm )=f_{\lambda}^* (\gm )\lambda ^{1/2}_G(r(\gm ))\otimes \lambda
^{1/2}_G(s(\gm )), \end{equation} where $f_{\lambda }^*(\gm
)=\overline{f_{\lambda }(\gm ^{-1})}$.

Let $x\in G_0$. The restriction of $\lambda _G$ to the submanifold
$G^x$ is a $1$-density, hence it has an associated measure on
$G^x$. As we identify the $1$-densities and their associated measures,
we use the same notation $\lambda^x$ for the restriction of
$\lambda_G$ to $G^x$ and the induced measure.  The equality $\gm \cdot
\lambda ^{s(\gm )}=\lambda ^{r(\gm )}$ proves that the $\lm^x$ form a
left Haar system; its is easy to check that each function (\ref{lHs})
is smooth.
 
It is easily checked that, under the correspondence $f\leftrightarrow
f_{\lm}$ defined in (\ref{identif}) and the replacement of $\lambda
_G$ by the left Haar system $(\lm^x)_{x\in G_0}$, the expressions
(\ref{lambda_convolution}) and (\ref{star}) are transformed into
(\ref{oldconv}) and (\ref{oldinv}), respectively. We leave the proof
of the isomorphisms of the completions of the \sta s in question to
the reader.
\end{proof}.
 \section{Strict deformation quantization}\label{Sdq} As we have seen,
 the idea of \ca ic deformation quantization is to relate a given
 Poisson algebra to a family of \ca s.
\begin{Definition}\ll{defpoisal}
A Poisson algebra is a complex commutative associative algebra
equipped with a Lie bracket $\{\, ,\,\}$ for the which the Leibniz
rule holds; that is, one has \begin{equation} \{f,gh\}=\{f,g\}h+
g\{f,h\} \label{leib}. \end{equation}

 A Poisson manifold is a manifold $P$ equipped with a Lie bracket
 $\{\, ,\,\}$ on $\cin(P)$ that together with pointwise multiplication
 turns $\cin(P)$ into a Poisson algebra.
\end{Definition}

This definition is due to Lichnerowicz \ci{Lic} and Kirillov
\ci{Kir}. The Lie bracket in a Poisson algebra is usually called the
Poisson bracket. For the theory of Poisson manifolds we refer to these
papers, and to the textbooks \ci{MR,Vai}.  A Poisson algebra is the
classical analogue of a \ca\ \ci{La3}.  This analogy is clear if one
sees a \ca\ as a non-associative version of a Poisson algebra, in
which the commutative product is given by the anti-commutator $xy+yx$
and the Lie bracket is $\{x,y\}=i(xy-yx)$. The Leibniz rule is then
satisfied. The analogy is even better when one defines a Poisson
algebra as a real algebra, and restricts the above two operations to
the self-adjoint part of a \ca.

In this paper, the key example of a Poisson algebra is provided by the
dual bundle $E^*$ of a Lie algebroid $E$ over $M$ (cf.\ Definition
\ref{defloid}). This Poisson structure was discovered in
\ci{coste,Cou}.  The corresponding bracket is most easily defined by
listing special cases by which it is uniquely determined; these are
\bea \label{pblieoid1} \{f,g\} & = & 0;  \\
\label{pblieoid2} \{\til{X},f\} & = &  (\rh\circ X) f;  \\
\label{pblieoid3} \{\til{X},\til{Y}\} & = & 
\wt{[X,Y]_{A(G)}}.  \eea Here $f,g\in\cin(M)$ are
regarded as functions on $E^*$ in the obvious way, and
$\til{X}\in\cin(E^*)$ is defined by a section $X\in\cin(M,E)$ through
$\til{X}(\theta)= \theta(X(\ta^*(\theta)))$.  See \ci{coste} for an
intrinsic definition.  In particular, a Lie groupoid $G$ canonically
determines a Poisson algebra $\cin(A^*(G))$.  This is the generic
Poisson algebra that we are going to quantize.

Our first definition of \ca ic deformation quantization uses the
concept of a continuous field of \ca s. We here state a reformulation
of Dixmier's familiar definition
\ci{Dix} due to Kirchberg--S.~Wassermann \ci{KW},
 which is tailor made for our applications.
\begin{Definition}\label{defcfca}
A continuous field of \ca s $(\GC,\{\A_{t},\phv_{t}\}_{t\in T})$ over
a locally compact Hausdorff space $T$ consists of a \ca\ $\GC$, a
collection of \ca s $\{\A_{t}\}_{t\in T}$, and a set
$\{\phv_{t}:\GC\raw\A_{t}\}_{t\in T}$ of surjective
$\mbox{}^*$-homomorphisms, such that for all $c\in\GC$
\begin{enumerate}
\item
the function $t\raw \| \phv_{t}(c)\|$ is in $C_0(T)$;
\item
one has $\| c \| =\sup_{t\in T}\| \phv_{t}(c)\|$;
\item
there is an element $fc\in\GC$ for any $f\in C_0(T)$ for which
$\phv_{t}(fc)=f(t)\phv_{t}(c)$ for all $t\in T$.
\end{enumerate}
\end{Definition}
The continuous cross-sections of the field in the sense of \ci{Dix}
consist of those elements $\{a_{t}\}_{t\in T}$ of $\prod_{t\in T} A_{t}$
for which there is a (necessarily unique)
 $a\in C$ such that $a_{t}=\phv_{t}(a)$ for all
$t\in T$.

Our first definition of \ca ic deformation quantization is now as
follows \ci{La3,La4}.
\begin{Definition}\label{defqua}
  A strict deformation
quantization of a Poisson manifold $P$ consists of
\begin{enumerate}
\item
a dense Poisson algebra $A^0\subset C_0(P)$ under the given Poisson
bracket $\{\, ,\,\}$ on $P$;
\item A subset $I\subseteq\R$ containing $0$ as an accumulation point;
\item 
a continuous field of \ca s $(\GC,\{\Ah,\phv_{\hbar}\}_{\hbar\in I})$,
with $\A_0=C_0(P)$;
\item
a linear map $\CQ:\Ao\raw\GC$ that satisfies (with
$\q(f)\equiv\phv_{\hbar}(\CQ(f))$) \bea \label{q0f} \CQ_0(f)& =& f; \\
\label{real} \q(f^*)& =&
\q(f)^* , \eea for all $f\in A^0$ and
$\hbar\in I$ , and for all $f,g\in A^0$ satisfies Dirac's condition
\begin{equation} \label{direq} \lho \|(i\hbar)\inv[\q(f),\q(g)]
-\q(\{f,g\})\|=0.  \end{equation}
\end{enumerate}
\end{Definition}
  In view of the comment after Definition \ref{defcfca}, for fixed
  $f\in A^0$ each family $\{\q(f)\}_{\hbar\in I}$ is a continuous
  cross-section of the continuous field in question. In view of
  (\ref{q0f}) this implies, in particular, that \begin{equation}
  \label{vneq} \lho \|\q(f)\q(g)-\q(fg)\| =0.  \end{equation} This
  shows that a strict deformation quantization yields asymptotic
  morphisms in the sense of $E$-theory \ci{connes}.  See \ci{La3} for
  an extensive discussion of quantization theory from the above
  perspective.

Every good definition is the hypothesis of a theorem. Indeed, our
first main result \ci{La3,La4} is as follows.
\begin{Theorem}\ll{main1}
Let $G$ be a Lie groupoid, with associated \ca s $C^*(G)$ and
$C_r^*(G)$, and Poisson manifold $A^*(G)$. There exists a strict
deformation quantization of $A^*(G)$ for which $I=\R$,
$A_0=C_0(A^*(G))$, and $A_{\hbar}=C^*(G)$ for $\hbar\neq 0$. In
particular, there exists a continuous field of \ca s over $\R$ with
these fibers. The same claim holds with $C^*(G)$ replaced by
$C_r^*(G)$.
\end{Theorem}

A different definition of \ca ic deformation quantization, which
generalizes Rieffel's original definition \ci{Ri2}, was introduced in
\ci{Ra1}. This definition is closer to the notion of formal
deformation quantization introduced in \ci{BFFLS} than Definition
\ref{defqua}.
\begin{Definition}\label{semi}
Let $P$ be a Poisson manifold. A semi-strict 
deformation quantization of $P$  consists of
\begin{enumerate}
\item
a dense Poisson algebra $A^0\subset C_0(P)$ under the given Poisson
bracket $\{\, ,\,\}$ on $P$;
\item A subset $I\subseteq\R$ containing $0$ as an accumulation point;
\item For each $h\in I$,  an associative  product $\times _{\hbar}$, an 
involution $^{*_{\hbar}}$, and a $C^*$-semi-norm $\| \cdot \|
_{\hbar}$ on $A^0$, such that
\begin{enumerate}
\item For $\hbar=0$ we recover the usual product, 
involution and norm of $C_0(P)$;
\item For each $f\in A^0$, the map $h\mapsto \| f\| _{\hbar}$ is continuous; 
\item For all $f,g\in A^0$ one has  
\begin{equation}
\label{diracram} \lho \| (i\hbar)\inv (f\times _{\hbar}g-g\times _{\hbar}f)
-\{ f,g\} \| _{\hbar} =0.
\end{equation}
\end{enumerate}
\end{enumerate}
\end{Definition}

A strict deformation quantization cannot necessarily be turned into a
semi-strict one, since $\CQ(A^0)\CQ(A^0)$ is not necessarily contained
in $\CQ(A^0)$, so that one may not be able to transfer the products on
the $\Ah$ to an $\hbar$-dependent product on $A^0$.  Conversely, a
semi-strict deformation quantization is not necessarily strict, for
the seminorms can fail to be norms. However, let $f\in A^0$ be a
non-null function. Then $\| f\| _0 \neq 0$ and by the condition 3.2 in
Definition \ref{semi} one has $\| f\| _{\hbar}\neq 0$ for $h$ in a
neighborhood of $0$. 

For each $\hbar\in I$, $N_{\hbar}=\left\{ f\in A^0| \ \| f\|
_{\hbar}=0\right\}$ is a two-sided closed ideal in $(A^0,\times
_{\hbar},{}^{*_{\hbar}},\| \cdot \| _{\hbar})$, and $A^0/N_{\hbar}$ is
a pre-$C^*$-algebra, with completion $A^-_{\hbar}$. Defining
$\CQ_{\hbar}$ as the canonical projection of $A^0$ onto
$A^0/N_{\hbar}$, we almost obtain a strict deformation quantization of
$P$; the only difficulty is that the $A^-_{\hbar}$ may not form a
continuous field of \ca s.

The counterpart of Theorem \ref{main1} for strict deformation
quantization is as follows \ci{Ra1}.
\begin{Theorem}\ll{main2}
Let $A(G)$ be the Lie algebroid of a Lie groupoid, with associated 
 Poisson manifold $A^*(G)$. There exists a semi-strict deformation quantization
of $A^*(G)$ over $I=\R$. 
\end{Theorem}

As explained above, this theorem does not follow from Theorem
\ref{main1}, though the proofs have many elements in common.  We now
briefly outline the organization of the proofs. Sections \ref{Cfg} and
\ref{tg} address the continuous field of \ca s claimed to exist in
Theorem \ref{main1}, and in similar vein contain the heart of the
proof of condition (b) of Definition \ref{semi} in Theorem
\ref{main2}.  Section \ref{Ft} develops the theory of the Fourier
transform on vector bundles.  In section \ref{Weyl} this theory is
used to define the map $\CQ$ of Theorem \ref{main1} as well as the
product $\times_{\hbar}$, the involution $*_{\hbar}$, and the seminorm
$\|\cdot\|_{\hbar}$ of Theorem \ref{main2}.  The proofs of both
theorems are then complete up to Dirac's condition (\ref{direq}) and
(\ref{diracram}).  Section \ref{lsL} develops local techniques that
give detailed information on the Poisson structure on
$A^*(G)$. Finally, (\ref{direq}) and (\ref{diracram}) are proved in
section \ref{ssc}.
\section{Continuous fields of groupoids}\label{Cfg}
 Rieffel has developed useful techniques for proving continuity of
 fields of $C^*$-algebras occurring in examples of strict deformation
 quantization in the context of groups \cite{Ri1}. The main result of
 this section, Theorem \ref{champs}, generalizes these results to the
 context of groupoids. This theorem, which is an application of
 results of Blanchard \cite{Bl}, gives information on the field of
 $C^*$-algebras associated to a continuous field of locally compact
 groupoids.  Although the examples studied in this paper concern Lie
 groupoids, we prove Theorem \ref{champs} for the general case of
 locally compact groupoids.

We first mention a known result. Recall that, when $E\subset G_0$,
$G_E$ is the subgroupoid of $G$ consisting of all $\gm\in G_1$ for
which $r(\gm)\in E$ and $s(\gm)\in E$.
\begin{Proposition}\label{suiteC*}
Let $G$ be a   locally compact groupoid with left Haar
system $(\lambda ^x)_{x\in G_0}$, and let $U$ be an open invariant subset of 
$G_0$.  Write $F=G_0\backslash U$.
\begin{enumerate}
\item The following sequence is exact:
\begin{equation}0\longrightarrow C^*(G_U)\stackrel{e}{\longrightarrow}C^*(G)
\stackrel{i^*}{\longrightarrow}C^*(G_F) \longrightarrow 0.
\end{equation}
 Here $e$ is firstly defined as map from $C_c(G_U)$ to $C_c(G_1)$ by
 extending a function on $G_U$ to one on $G_1$ by making it zero on
 the complement of $G_U$, and secondly extended to a map from
 $C^*(G_U)$ to $C^*(G)$ by continuity.  Similarly, $i^*$ is firstly
 defined from $C_c(G_1)$ to $C_c(G_F)$ as the pullback of the
 inclusion $G_F\hookrightarrow G_1$, and then extended by continuity.
\item If the groupoid $G_F$ is amenable, then the following sequence is exact: 
 \begin{equation}
0\longrightarrow C^*_r(G_U)\stackrel{e}{\longrightarrow}C^*_r(G)
\stackrel{i^*}{\longrightarrow}C^*_r(G_F)\longrightarrow 0.
\end{equation}
\end{enumerate}
\end{Proposition}

This result was given by Torpe \cite{torpe} in the case 
of $C^*$-algebras associated to foliations, and by Hilsum and Skandalis
\cite{hilsum} in the general case; also see \cite{Re1}, p.102. 
A complete proof, based on Renault's theorem of disintegration of
representations \cite{Re2}, may be found in \cite{Ra1}.  A
counterexample that shows that the second sequence of reduced
$C^*$-algebras is not necessarily exact for non-amenable $G_F$ is
given in \cite{Re3}.

\begin{Definition}
A field of groupoids is a triple $(G,T,p)$, with $G$ a groupoid, $T$ 
a set, and $p:G\rightarrow T$ a surjective map such that $p=p_0\circ r=p_0
\circ s$, where $p_0=p_{|_{G_0}}$. If $G$ is  locally compact,
$T$  Hausdorff, and $p$  continuous and open,  we say that
$(G,T,p)$ is a continuous field of locally compact groupoids. If $G$ is a 
Lie groupoid, $T$ a manifold, and $p$  a submersion, $(G,T,p)$ is called a
smooth field of Lie groupoids.
\end{Definition}

If $(G,T,p)$ is a field of groupoids and $Y\subset T$, then $A=p_0^{-1}(Y)$ 
is an invariant subset of $G_0$, and $G_A=p^{-1}(Y)$ is a subgroupoid of 
$G$. In the case of a continuous field of locally compact groupoids, 
 \begin{equation}
G(t)=p^{-1}(\lbrace t\rbrace )
\end{equation} is a closed locally compact  
subgroupoid of $G$ for every 
$t\in T$. In the case of a smooth field of Lie groupoids, 
$G(t)$ is a Lie groupoid for every $t\in T$.

In this section, $(G,T,p)$  is  a continuous field of   locally compact 
groupoids over a locally compact Hausdorff space $T$, for which there exists
a left Haar system $(\lambda ^x)_{x\in G_0}$ on $G$. 

Let $C_0(T)$ be the $C^*$-algebra of continuous functions on $T$ that
vanish at infinity. We define a structure of $C_0(T)$ left module on
the space of continuous compactly supported functions $C_c(G_1)$ by
$(fa)(\gm )=f(p(\gm ))a (\gm )$, where $f\in C_0(T)$ and $a\in
C_c(G_1)$. For every $f\in C_0(T)$ and $a,b\in C_c(G_1)$, the
following equalities hold:
\bea
f(a*b) & =  & (fa)*b=a*(fb); \nn\\
(fa)^* & = & f^*a^*.
\eea
We also have $C_0(T)C_c(G_1)=C_c(G_1)$. Indeed, for every $a\in
C_c(G_1)$, there is a function $f\in C_c(T)$ such that $f=1$ on
$p(\mbox{supp}\, a)$, and then $a=fa\in C_0(T)C_c(G_1)$.

An easy rewriting of Lemma 1.13 from \cite{Re1} proves

\begin{Lemma}\label{lema}
For every representation $\pi:C_c(G_1)\rightarrow {\mathcal
L}(\mathcal{H})$ that is continuous for the inductive limit topology
there is a unique continuous representation $\phi:C_0(T)\rightarrow
{\mathcal L}(\mathcal{H})$ such that $\pi(fa)=\phi(f)\pi(a)$.
\end{Lemma}

Using the previous lemma for $f\in C_0(T)$, $a\in C_c(G_1)$, and $\pi$
continuous representation of $C_c(G_1)$, we have $\left\|
\pi(fa)\right\| \leq \left\|\phi(f)\right\| \left\| \pi(a)\right\|
\leq
\| f\| \| a\|$, hence $\| fa\| \leq \| f\| \| a\|$. 
For every $f\in C_0(T)$, the map $C_c(G_1)\ni a\mapsto fa\in C_c(G_1)$ 
has a continuous extension on $C^*(G)$. This provides a
$C_0(T)$ Banach module structure on $C^*(G)$. 

Corollary 1.9 in \cite{Bl} shows that $C_0(T)C^*(G)$ is closed in
$C^*(G )$. But one has $C_0(T)C^*(G)\supset C_0(T)C_c(G_1)=C_c(G_1)$,
and we obtain that $C^*(G)$ is a nondegenerate module. For every
$a,b\in C^*(G)$ and every $f\in C_0(T)$, the conditions
$f(a*b)=(fa)*b=a*(fb)$ and $(fa)^*=f^*a^*$ are easily verified, hence
$C^*(G)$ is a $C_0(T)$ $\mbox{}^*$-algebra (cf.\ \cite{Bl}). Similar
arguments prove that $C^*_r(G)$ is a $C_0(T)$ $\mbox{}^*$-algebra.

\begin{Lemma}\label{ouverte}
Let $T$, $Y$ be topological spaces, $p:Y\rightarrow T$ an open onto
map, and $\varphi :Y\rightarrow \R$ an upper semicontinuous function
such that for every $t\in T$ one has sup$\left\{ \varphi
(y)|p(y)=t\right\} <\infty$. Then the function $\psi :T\rightarrow \R$
given by $\psi (t)=\stackunder{p(y)=t}{\sup}
\varphi (y)$ is lower semi-continuous.
\end{Lemma}

\begin{proof}
Let $t_i$ be a net converging in $T$ to $t$, and pick $a\in \R$. We
show that $\psi (t_i)\leq a$ implies $\psi (t)\leq a$. Let $y\in Y$ be
such that $p(y)=t$. A lemma on open maps (cf.\ \cite{DF}, p.126)
proves the existence of a net $y_i\in Y$, for which $p(y_i)=t_i$ and
$y_i\rightarrow y$.  For every $i$, $\varphi (y_i)\leq \psi (t_i)\leq
a$. But $\varphi$ is lower semicontinuous, so $\varphi (y)\leq a$ and
$\psi (t)\leq a$.
\end{proof}

We now come to the main result of this section.
\begin{Theorem}\label{champs}
Let $(G,T,p)$ be a continuous field of locally compact groupoids, take
$a\in C_c(G_1)$, and, for each $t\in T$, write $a_t=a_{|_{G(t)}}$. Then 
\begin{enumerate}
\item
The map $t\mapsto \left\| a_t\right\| _{C^*(G(t))}$ is upper
semicontinuous.
\item
 The map $t\mapsto \left\| a_t\right\| _{C^*_r(G(t))}$ is lower 
semicontinuous.
\end{enumerate}
\end{Theorem}

\begin{proof}
1. For $t\in T$, $C_t(T)=\left\{ f\in C_0(T)|f(t)=0\right\}$ is a closed
two sided ideal, hence $C_t(T)C^*(G)$ is a closed two sided ideal 
in $C^*(G)$. Corollary 1.9 in  \cite{Bl} applied to the
$C_t(T)$ Banach module $C^*(G)$ shows that $C_t(T)C^*(G)$ is closed in 
$C^*(G)$. Let $\pi _t :C^*(G)\rightarrow C^*(G)/C_t(T)C^*(G)$ be the quotient 
map. By Lemma 1.10 from \cite{Bl} one has $\left\| \pi _t(a)\right\| =
\stackunder{f\in C_0(T)}{\inf}\left\| (1-f+f(t))a\right\|$;
hence the map $t\mapsto \left\| \pi _t(a)\right\|$ is upper 
semicontinuous as the infimum of a family of continuous functions.

Denote  the closed invariant subset $p_0^{-1}(\left\{ t\right\})$ of 
$G_0$ by $F$, and let $U=G_0\backslash F$. Obviously, $G(t)=G_F$. 
We claim that
\begin{equation}\label{i}
i\left( C_c(G_U)\right) =C_t(T)C_c(G),
\end{equation}
where $i:C^*(G_U)\rightarrow C^*(G)$ is as in Proposition \ref{suiteC*}. 
Let $a\in i\left( C_c(G_U)\right)$. In $T$ there is an open neighborhood $V$ 
of the compact set $p(\mbox{supp}\, a)$ such that $t\notin V$. We take 
$f\in C_c(T)$ with $f=1$ on $p(\mbox{supp}\, a)$ and $\mbox{supp}\, 
f\subset V$, and then $a=fa\in C_t(T)C_c(G_1)$. The other inclusion is obvious.

Taking the norm closure in (\ref{i}), we obtain $i(C^*(G_U))=C_t(T)C^*(G)$.
With Proposition \ref{suiteC*} this shows that $C^*(G(t))=C^*(G)/C_t(T)
C^*(G)$, so $\left\| \pi _t(a)\right\| =\left\| a_t\right\|$, which ends the 
proof.
 
2. The left regular representation $\pi^L_x(f):L^2(G_1,\lambda
^x)\rightarrow L^2(G_1,\lambda ^x)$ of $C_c(G_1)$ associated to $x\in
G_0$ is given by
\begin{equation}
\pi^L_x(f)\xi (\gm )=\dint_{G^x} f(\gm \gm _1)\xi (\gm _1^{-1})d\lambda ^x
(\gm _1).
\end{equation}
 Set $\| \xi \| _{\infty} =\stackunder{x\in G_0}{\sup}
\| \xi \| _{L^2(G_1,\lambda ^x)}$; then
$$\left\| \pi^L_x(f)\right\| =\sup\left\{ \left\langle \pi^L_x(f)\xi ,\eta 
\right\rangle | \xi ,\eta \in C_c(G_1), \| 
\xi \| _{\infty} <\infty ,\| \eta \| _{\infty} <\infty \right\}.$$
The map $x\mapsto \left\langle \pi^L_x(f)\xi ,\eta \right\rangle$ is
continuous for each $f\in C_c(G_1)$ and $\xi ,\eta \in C_c(G_1)$,
hence $x\mapsto \left\| \pi^L_x(f)
\right\|$ is lower semicontinuous. But $\| a_t\| =\stackunder{t\in p^{-1}(t)}
{\sup} \| \pi_x^L(a_t)\|$, and Lemma \ref{ouverte} ends the proof.
\end{proof}

\begin{Corollary}\label{champcontmoy}
If $(G,T,p)$ is a continuous field of locally compact groupoids and 
$a\in C_c(G_1)$, then
\begin{enumerate}
\item
The maps $t\mapsto \| a_t\|$ and $t\mapsto \| a_t\| _r$
 are continuous at every $t\in T$ for which the groupoid $G(t)$ is 
amenable.
\item
If $G$ is amenable, then the maps $t\mapsto \| a_t\|$ 
and $t\mapsto \| a_t\| _r$ are continuous.
\end{enumerate}
\end{Corollary}

\begin{proof}
1. Straightforward.

2. By Proposition 5.2.3 from \cite{AR}, $G$ is amenable if and 
only if $G(t)$ is amenable for every $t\in T$.
\end{proof}
\section{The tangent groupoid}\label{tg}
The tangent groupoid of a manifold was introduced by Connes
\ci{connes}.  This idea was generalized to arbitrary Lie groupoids by
Hilsum and Skandalis
\ci{hilsum} under the name normal groupoid,
 and by Weinstein \ci{Wei89} under the name blowup.  Connes's
 construction corresponds to the pair groupoid of a manifold. We here
 use the name ``tangent groupoid'' also for the general case.
\begin{Definition}\ll{defnormoid} 
Let $G$ be a Lie groupoid with Lie algebroid
$A(G)$.  The tangent groupoid $\hat{G}$ is a Lie groupoid with base
$\hat{G}_0=[0,1]\x G_0$, defined by the following structures.
\begin{itemize}
\item
As a set, $\hat{G}_1=A(G)\cup \{(0,1]\x G_1\}$. We write elements of
$\hat{G}$ as pairs $(\hbar,u)$, where $u\in A(G)$ for $\hbar=0$ and
$u\in G$ for $\hbar\neq 0$.  Thus $A(G)$ is identified with $\{0\}\x
A(G)$.  \item As a groupoid, $\hat{G}=\{0\x A(G)\} \cup \{ (0,1]\x G\}$.
Here $A(G)$ is regarded as a Lie groupoid over $G_0$ with $s=r=\ta$
(the bundle projection of $A(G)$), and addition in the fibers as the
groupoid multiplication.  The groupoid operations in $(0,1]\x G$ are
those in $G$. For example, for $\hbar\neq 0$ one has
\begin{equation}
\hat{r}(\hbar,u)=(\hbar,r(u).\ll{rtilde}
\end{equation}
\item
The smooth structure on $\hat{G}_1$, making it a manifold with boundary,
is as follows.  To start, the open subset $\CO_1=(0,1]\x G_1\subset
\hat{G}_1$ inherits the product manifold structure.  Let $G_0\subset
 V\subset A(G)$ and $\io(G_0)\subset W\subset G_1$, as in
Theorem \ref{tntoid}. Let $\CO$ be the open subset of $[0,1]\x A(G)$
(equipped with the product manifold structure; this is a manifold with
boundary, since $[0,1]$ is), defined as $\CO=\{(\hbar,X)\,|\,\hbar
X\in  V\}$.  Note that $\{0\}\x A(G)\subset \CO$.  The map
$\ps:\CO\raw \hat{G}_1$ is defined by \bea \ps(0,X) & = & (0,X); 
\ll{OX}\nn \\
\ps(\hbar,X) & = & (\hbar, \Exp^W(\hbar X)). 
\ll{Whbar} \eea Since $\Exp^W:
 V\raw W$ is a diffeomorphism (cf.\ Proposition
\ref{tntoid}) we see that $\ps$ is a bijection from $\CO$ to
$\CO_2=\{0\x A(G)\} \cup \{(0,1]\x W\}$.  This defines the
smooth structure on $\CO_2$ in terms of the smooth structure on $\CO$.
Since $\CO_1$ and $\CO_2$ cover $\hat{G}_1$, this specifies a smooth
structure on $\hat{G}_1$.
\end{itemize} 
\end{Definition}

Using $\Exp^L$ rather than $\Exp^W$ in this definition, one obtains
the smooth structure defined in \ci{hilsum}, which is equivalent to
the one constructed above.  A proof that the change of coordinates
from $\CO_1$ to $\CO_2$ on their intersection is smooth may be found
in \ci{hilsum}.

We omit the proof that $\hat{G}$ is a Lie groupoid; see \ci{hilsum}, and, for
full details, \ci{Ra1}. It follows that 
 $\hat{G}$ is a smooth field of Lie groupoids
over $\R$.
\begin{Proposition}\label{2suites}
Let $G$ be a Lie groupoid with tangent groupoid $\hat{G}$.
Writing $\R^*=\R\backslash \{0\}$, the following  sequences are exact:
\begin{eqnarray}
0 & \rightarrow & C_0(\R ^*)\otimes C^*(G)\rightarrow
C^*(\hat{G})\rightarrow C_0(A^*(G))\rightarrow 0; \\ 0 & \rightarrow
& C_0(\R ^*)\otimes C^*_r(G)\rightarrow C^*_r(\hat{G})
\rightarrow C_0(A^*(G))\rightarrow 0.
\end{eqnarray}
\end{Proposition}

\begin{proof}
Consider the open invariant subset $U=G_0\times \R ^*$ of
$\hat{G}_0$, with complement $F=G_0\backslash U$. Then $\hat{G}
_U=G\times \R ^*$ and $\hat{G} _F= A(G) \times \left\{
0\right\}$. By Proposition \ref{suiteC*} we have the exact sequence
$$0\rightarrow C^*(G\times \R ^*)\rightarrow C^*(\hat{G})
\rightarrow C^*(A(G))\rightarrow 0.$$
 Gelfand's theorem implies $C^*(A(G))=C_0(A^*(G))$ (also cf.\ section
 \ref{Ft}), and since $C^*(G\times
\R ^*)$  is the $C^*$-algebra of the trivial field of the \ca s $C^*(G)$ 
on $\R^*$, we have $C^*(G\times \R ^*)=C_0(\R ^*)\otimes C^*(G)$.

The second exact sequence is a consequence of Proposition
\ref{suiteC*} as well, since $A(G)$ is commutative as a groupoid 
(cf.\ Definition \ref{defnormoid}), 
and therefore amenable \ci{AR}.
\end{proof}
\begin{Corollary} (cf.\ \cite{connes})
Let $M$ be a manifold. Then the tangent groupoid
 $\widehat{M\x M}=(M\times M\times \R ^*)
\bigcup (TM \times\left\{ 0\right\})$ 
is amenable, and the  sequence
\begin{equation}
0\rightarrow C_0(\R ^*)\otimes {\mathcal K}(L^2(M))\rightarrow
C^*(\widehat{M\x M})
\rightarrow C_0(T^*M)\rightarrow 0
\end{equation}
is exact.
\end{Corollary}
\begin{proof}
By Proposition 5.2.3 in \cite{AR}, a groupoid bundle is amenable if
and only if all its fibers are amenable. Since the Lie algebroid $A(G)$
is amenable as a groupoid, the amenability of $\hat G$ is reduced to
the amenability of the groupoid $G$. In this case, $C^*(\hat
G)=C^*_r(\hat G)$, cf.\
\cite{Re1,AR}. Now use Proposition \ref{2suites}
for the amenable groupoid $G=M\times M$ to finish the proof.
\end{proof}
Recall Definition \ref{defcfca}, whose notation we adopt.
\begin{Theorem}\label{323}
Let $G$ be a Lie groupoid with tangent groupoid $\hat{G}$. 
For $\hbar\in\R$, define 
\bea
G(0) & = & A(G); \nn \\
G(\hbar) & = & G  \:\:\forall \hbar\neq 0. \ll{Ghbar}
\eea
Define $\hat{\phv}_{\hbar}:\cci(\hat{G})\raw \cci(G(\hbar))$ as the
pullback of the inclusion $G(\hbar)\hookrightarrow \hat{G}$ (cf.\
Definition
\ref{defnormoid}); in other words,
 $\hat{\phv}_{\hbar}(\hat{f})$ is the restriction of $\hat{f}\in
\cci(\hat{G})$ to $G(\hbar)$. 
\begin{enumerate}
\item
Each $\hat{\phv}_{\hbar}$ may be extended by continuity to a
surjective $\mbox{}^*$-homomorphism $\phv_{\hbar}:C^*(\hat{G})\raw
C^*(G(\hbar))$, and also to a surjective $\mbox{}^*$-homomorphism
$\phv_{(r)\hbar}:C_r^*(\hat{G})\raw C_r^*(G(\hbar))$.
\item
The \ca s $C= C^*(\hat{G})$ and  $A_{\hbar}=C^*(G(\hbar))$, and the maps 
$\phv_{\hbar}$ form a continuous field of \ca s over $I=\R$.
\item
The \ca s $C=C_r^*(\hat{G})$ and $A_{\hbar}=C_r^*(G(\hbar))$, and the maps 
$\phv_{(r)\hbar}$ form a continuous field of \ca s over $I=\R$.
\end{enumerate}
\end{Theorem}
\begin{proof}
\begin{enumerate}
\item Using the definition of the norms in question, one checks that each map
$\hat{\phv}_{\hbar}$ is contractive. The $\mbox{}^*$-homomorphism property is
obvious on $\cci(\hat{G})$ from the definition of $\hat{G}$ and $C^*(G)$, and 
extends by continuity.
\item Condition 1 in Definition \ref{defcfca} follows at $\hbar\neq 0$
since the field of $C^*$-algebras is trivial away from 0. At $\hbar=0$
continuity follows from Corollary 
\ref{champcontmoy}, since $A(G)$, being a commutative groupoid, 
is amenable \ci{AR}.
\item The reduced case is proved in the same way.
\end{enumerate}
\end{proof}

For a different proof of this theorem, note that each family
$\hat{\phv}_{\hbar}(\hat{f})$, $\hat{f}\in \cci(\hat G)$, is stable
under convolution and involution, and dense in $C^*_{(r)}(G)$. One
then uses Prop.\ 10.3.2 in \ci{Dix} and the equivalence between
Kirchberg--S. Wassermann's Definition \ref{defcfca} and Dixmier's
definition Def.\ 10.3.1 in \ci{Dix} of a continuous field of \ca
s. Condition 10.1.2 (ii) in \ci{Dix} is then proved in the same way as
above.

A different proof of Theorem \ref{323}, which makes no use of the
results in section \ref{Cfg} is possible as well \ci{La3,La4}. This
proof is based on the following lemma, due Lee; cf.\ Theorem 4 in
\ci{Lee}.
\begin{Lemma}\ll{tomiyama}
Let $C$ be a \ca, and let $\ps:\Pri(C)\raw T$ be a continuous and
open map from the primitive spectrum $\Pri(C)$ (equipped with the
Jacobson topology \ci{Dix}) to a locally compact Hausdorff space $T$.
Define $I_{t}=\cap \ps\inv(t)$; i.e., $c\in I_{t}$ iff
$\pi_{I}(c)=0$ for all $I\in\ps\inv(t)$ (here $\pi_{I}(C)$ is
the irreducible \rep\ whose kernel is $I$).  Note that $I_t$ is a
closed two-sided ideal in $C$.

Taking $A_t= C/I_t$ and $\phv_t:C\raw A_t$ to be the canonical
projection, $(C,\{A_t,\phv_t\}_{t\in T})$ is a continuous field of
\ca s.
\end{Lemma}

For a proof see \ci{Lee} or \ci{ENN1}.  We apply this lemma with
$C=C^*(\hat{G})$ and $T=I=\R$. In order to verify the assumption in
the lemma, we first note that $I_0\simeq C_0((0,1])\ot C^*(G)$, as
follows from a glance at the topology of $\hat{G}$. Hence
$\Pri(I_0)=(0,1]\x \Pri(C^*(G))$, with the product topology.
Furthermore, one has $C^*(\hat{G})/I_0\simeq C^*(A(G))\simeq
C_0(A^*(G))$.  Hence $\Pri(C^*(\hat{G})/I_0)\simeq A^*(G)$.  Using this
in Prop.\ 3.2.1 in \ci{Dix}, with $A=C^*(\hat{G})$ and $I$ the ideal
$I_0$ generated by those $f\in\cci(\hat{G})$ that vanish at $\hbar=0$,
yields the decomposition \begin{equation} \Pri(C^*(\hat{G}))\simeq
A^*(G)\cup \{(0,1]\x \Pri(C^*(G))\}, \ll{decspgroid} \end{equation} in which
$A^*(G)$ is closed.  This does not provide the full topology on
$\Pri(C^*(\hat{G}))$, but it is sufficient to know that $A^*(G)$ is not
open. If it were, $(0,1]\x \Pri(C^*(G))$ would be closed in
$\Pri(C^*(\hat{G}))$, and this possibility can  be excluded
by looking at the topology of $\hat{G}$ and the definition of the
Jacobson topology.

Using (\ref{decspgroid}), we can define a map $\ps:\Pri(C^*(\hat{G}))
\raw [0,1]$ by $\ps(I)=0$ for all $I\in A^*(G)$ and
$\ps(\hbar,I)=\hbar$ for $\hbar\neq 0$ and $I\in
\Pri(C^*(G))$.  It is clear from the preceding considerations that
$\ps$ is continuous and open. Using this in Lemma \ref{tomiyama}, one
sees that $I_{\hbar}$ is the ideal in $C^*(\hat{G})$ generated by
those $\hat{f}\in\cci(\hat{G})$  vanish at $\hbar$. Hence $A_0\simeq
C_0(A^*(G))$, as above, and $A_{\hbar}\simeq C^*(G)$ for $\hbar\neq
0$. Theorem \ref{323} then follows from Lemma \ref{tomiyama}. 

We now give a left Haar system for the tangent groupoid $\hat{G}$.
Firstly, pick a positive section $\mu \in C^{\infty}
(G_0,|\Om |^1(A(G) ))$, and let $\lambda$ be the associated smooth, 
positive, left invariant section of $|\Om |^1(\ker Tr)$, as in 
Proposition \ref{isodensit}; hence $\mu=\lm|_{G_0}$. 
Now define a section $\hat{\lm}\in \cin(\hat{G}_1,|\Om |^1(\ker  T\hat{r})$
by
\bea
\hat{\lambda}(x,X,0) & = & \lambda (x); \nn \\
 \hat{\lambda}(\gm ,\hbar) & = &  |\hbar|^{-p}\lambda (\gm). \ll{tillam}
\eea
Here $p$ is the dimension of the typical fiber of $A(G)$; the factor
$|\hbar|^{-p}$ is necessary in order to have a smooth system also at
$\hbar=0$, as is easily verified using the manifold structure on
$\hat{G}$.  This section is smooth, positive and left invariant, so
that it defines a left Haar system
$(\hat{\lm}^{(\hbar,x)})_{(\hbar,x)\in\hat{G}_0}$ for $\hat{G}$ by
Proposition
\ref{identification}. 
The $\mbox{}^*$-algebraic structure on $\cci(\hat{G},\hat{\lm})$ defined by
(\ref{oldconv})  and (\ref{oldinv}) with (\ref{tillam})
becomes
 \bea \hat{f}*\hat{g}(0,\xi_x) & = &
\int_{\ta\inv(x)} \hat{f}(0,\xi-\et_x)\hat{g}(0,\et_x) d\mu^x(\et_x);
 \\ \hat{f}*\hat{g}(\hbar,\gm) & = & |\hbar|^{-p}
\int_{G^{s(\gm)}}  \hat{f}(\hbar,\gm\gm_1)\hat{g}(\hbar,\gm_1\inv) 
d\lm^{s(\gm)}(\gm_1); 
\ll{eq1} \\ \hat{f}^*(0,\xi) & = & \ovl{\hat{f}(0,-\xi)} ; 
\\ \hat{f}^*(\hbar,\gm) & = &
\ovl{\hat{f}(\hbar,\gm\inv)}.  \ll{longea} \eea
Here $\xi_x\in\ta\inv(x)$, and $\mu^x$ is the measure on
$\ta\inv(x)\subset A(G)$ generated by the density $\mu(x)$ (cf.\
(\ref{calculintegral})). Finally, $(\lm^x)$ is the left Haar system on
$G$ defined by $\lm$ according to Proposition
\ref{identification}. These formulae should be compared with (\ref{oldconv})
 and (\ref{oldinv}).
\section{The Fourier transform on vector bundles}\label{Ft}
In this section we extend the notion of a Fourier transform from
$\R^n$ to vector bundles, and show that it remains an isomorphism of a
suitably defined Schwartz space of rapidly decreasing functions.

Let $M$ be a manifold of dimension $n$, and let $E$ be a vector bundle
over $M$ with fiber dimension $p$.  Now $E$ is a commutative Lie
groupoid if the source and range projections are both equal to the
bundle projection, and groupoid multiplication is addition in each
fiber. The $C^*$-algebra $C^*(E)$ is then commutative, too. Using the
proposition on page 582 of \cite{DF}, it can be shown that the maximal
ideal space of $C^*(E)$ is $E^*$. The Gelfand transform is an
isomorphism between the $C^*$-algebras $C^*(E)$ and $C_0(E^*)$. Its
explicit form is a fiberwise Fourier transform ${\mathcal
F}:C^*(E)\rightarrow C_0(E^*)$. On the convolution algebra
$C_c^{\infty}(E,{\frak E})$ the Fourier transform is given, for
$\theta_x\in E_x^*\subset E^*$, by $$({\mathcal F}\rho
)(\theta_x)=\stackunder{E_x}{\dint}e^{- i
\langle \theta_x,\xi _x\rangle} \rho (\xi _x).$$

It is easier to write formulas for functions rather than densities. To
do so, fix a positive 1-density $\mu \in C^{\infty}(M,|\Om |^1(E))$;
cf.\ the proof of Proposition \ref{identification}. For every $x\in M$,
the density $\mu (x)\in |\Om |^1(E_x)$ defines a translation invariant
measure $\mu^x$ on $E_x$, and the family $(\mu^x)_{x\in M}$ is a
smooth Haar system for $E$. The Fourier transform is denoted by
${\mathcal F}_{\mu}:C^*(E,\mu )
\rightarrow C_0(E^*)$ in order to emphasize its dependence on $\mu$. For
$f\in L^1(E,d\mu^x)$, $\theta_x\in E_x^*$, we then have 
\begin{equation} \label{fourierlocal}
({\mathcal F}_{\mu}f)(\theta_x)=\dint f(\xi _x)e^{- i\langle \theta_x,
\xi _x\rangle}d\mu^x(\xi _x).
\end{equation}

We  now generalize  the definition given  by Rieffel in the case of the 
trivial bundle $M\times \R^p$ (cf.\ \cite{Ri3}) to define a Schwartz space
on $E$. But first  we fix some notations.

The variables in $\R ^n\times \R ^p$  are  denoted $(u,v)$, and for 
$\beta \in \N ^p$ we  put $\d ^{\beta}_vF=\dfrac{\d ^{\beta _1+ 
\cdots +\beta _p}F}{\d 
v_1^{\beta _1} \cdots \d v_p^{\beta _p}}$. The Fourier transform
${\mathcal F}_u$ on $\R ^n\times
\R ^p$ with $u$ constant  is given by 
$({\mathcal F}_uF)(u,w)=
\dint F(u,v)e^{- i\langle v,w\rangle} dv$.

Let $(e_1,\ldots,e_p)$ be a local frame of $E$, with dual frame
$(e^*_1,\ldots,e^*_p)$ of $E^*$, let $(q,\lambda )$ be local
coordinates on $E$, and finally let $(q,\epsilon )$ be local
coordinates of $E^*$. The expression of a function $f$ on $E$ in local
coordinates is denoted by the corresponding capital letter $F$. Then
(\ref{calculintegral}) for $f\in L^1(E,\mu^x)$ becomes
\begin{equation}
\int f\left(\stackrel{n}{\stackunder{i=1}{\sum}}
v_ie_i(x)\right) \mu _e (x)dv=\int fd\mu^x .
\end{equation}
\begin{Definition}
We say that a function $f:E\rightarrow \C$ is $M$-compactly supported
when $\pi (\mbox{supp}\, f)$ is relatively compact in $M$.  The set of
continuous $M$-compactly supported functions is called $C_{c,M}(E)$.

A function $f\in C_{c,M}(E)$ is said to be rapidly decreasing if
$\lambda ^{\alpha}f(q,\lambda )$ is bounded for every $\alpha \in \N
^p$, where $\lambda ^{\alpha}= \lambda _1^{\alpha _1}\ldots \lambda
_p^{\alpha _p}$, with $\lambda _1,\ldots,\lambda _p$ the coordinates
of $\lambda$ in the frame $(e_1,
\ldots,e_p)$.
\end{Definition}

This definition is independent of the frame. Indeed, let
$(e_1,\ldots,e_p)$ be a frame of $E$ on an open subset $U$ of $M$, and
suppose that $\lambda ^{\alpha}f(q,\lambda )$ is bounded for every
$\alpha \in \N ^p$.  If $(e_1^{\prime},\ldots,e_p^{\prime})$ is
another frame of $E$ on $U$, let $g:U
\rightarrow GL(p,\R )$ be the function given by $e^{\prime}=eg$. If we write
the same point of $E$ in the two frames, its coordinates are related by 
$\lambda _i^{\prime}=
\stackunder{j}{\sum}\lambda _jg_{ji}$. To show that ${\lambda ^{\prime}}^{
\alpha}f(q,\lambda ^{\prime})$ is bounded, remark that expanding
${\lambda _1^{\prime}}^{\alpha _1}\ldots{\lambda _p^{\prime}}^{\alpha _p}$ we 
obtain a polynomial in $\lambda _1,\ldots,\lambda _p$ with 
continuous functions on $M$ that vanishes outside the compact closure of
$\pi (\mbox{supp}f)$ as  coefficients.
\begin{Definition}
We say that a function $f\in C^{\infty}(E)$ is of Schwartz type on $E$ if $f$
is $M$-compactly supported and 
$\d ^{\beta}_{\lambda}f=\dfrac{\d ^{\beta _1+\cdots+\beta _p}f}{\d \lambda _1^{
\beta _1}\cdots\lambda _p^{\beta _p}}$ is rapidly decreasing 
for every $\beta =(\beta _1,\ldots,\beta _p)\in \N ^p$. 
The set of Schwartz functions on $E$  is denoted by $S(E)$.
\end{Definition}
\begin{Proposition}\label{subalgebra}
$S(E)$ is a dense $\mbox{}^*$-subalgebra of $C^*(E,\mu )$ and $S(E^*)$
is a dense $\mbox{}^*$-subalgebra of $C_{0}(E^{*})$.
\end{Proposition}
\begin{proof}
We show that $S(E)$ is closed under convolution. For $f,g$ in $S(E)$
it is obvious that $f\ast g$ is $M$-compactly supported. Using a
partition of unity argument, it easily follows that $f\ast g\in
S(E)$. It is trivial that $S(E)$ is closed under involution.  Now
$S(E)$ contains the smooth compactly supported functions on $E$, and
by a result from \cite{DF}, page 140, $C_{c}^{\infty }(E)$ is dense in
$C^*(E,\mu )$; it follows that $S(E)$ is dense in $C^*(E,\mu )$.

For the second part of the proposition, $S(E^{*})$ is obviously a 
$\mbox{}^*$-subalgebra of $C_{0}(E^{*})$  containing the 
dense subalgebra $C_{c}^{\infty }(E^{*})$ of $C_0(E^*)$.
\end{proof}
\begin{Lemma}
If $f\in S(E)$, then ${\mathcal F}_{\mu}f\in S(E^*)$
\end{Lemma}
\begin{proof}
The restriction of $f$ to $E_x$ is integrable because it is rapidly
decreasing, so (\ref{fourierlocal}) makes sense in this case. It is
then easy to show that ${\mathcal F}_{\mu}f$ is smooth and
$M$-compactly supported.

Choosing a suitable partition of unity, we may assume that the
projection onto $M$ of the support of $f$ is a subset of the domain of
a chart $\alpha :U\rightarrow \R ^n$ of $M$. We can also suppose that
there exists a frame $(e_1,\ldots,e_p)$ of $E$ on $U$.

The expression of $f$ in the corresponding local coordinates of $E$ is 
given by $F:\alpha (U)\times \R ^p\rightarrow \C$, $F(u,v)=f
(\sum v_ie_i (\alpha ^{-1}(u)))$. It is straightforward to see that
the local expression of ${\mathcal F}_{\mu}f$, denoted by $\hat{F}$, satisfies
$\hat{F}(u,w)  = \mu _e(\alpha ^{-1}(u))({\mathcal F}_uF)(u,w)$.
The relation $w^{\alpha} \d ^{\beta}_w\hat{F}(u,w)  =  
\mu_e(\alpha ^{-1}(u))(-i)^{|\alpha |+|\beta |}
{\mathcal F}_u\left( \d ^{\alpha}_v(v^{\beta}F)\right) (u,w)$ reduces the 
problem to the fact that ${\mathcal F}_uG$ is bounded for every $G$ rapidly 
decreasing on the trivial bundle $\R^n\times\R^p$ of base $\R^n$. 

But $\left| {\mathcal F}_uG(u,w)\right| \leq \dint |F(u,v)|\,dv$, and the 
proof is finished by the remark that on the right side we have a continuous 
compactly supported function in $u$.
\end{proof}

For a vector space $V$ and a nowhere vanishing element $\rho$ of $|\Om
|^{\alpha}(V)$, let $\rho^{*}$ be the element of $|\Om
|^{\alpha}(V^*)$ that satisfies $\rho^{*}(v_1^*\wedge \cdots \wedge
v_n^*)= \dfrac{1}{\rho (v_1\wedge \cdots \wedge v_n)}$ for every pair
of dual bases $\left\{ v_1,\ldots,v_n\right\}$ on $V$ and $\left\{
v_1^*,\ldots,v_n^*\right\}$ on $V^*$.  Then $\mu_{*}(x)=\left( \mu
(x)\right)^{*}$ defines a 1-density $\mu_{*}\in C^{\infty}(M,|\Om
|^1(E^*))$, and hence a left Haar system $(\mu_*^x)_{x\in M}$ on $E^*$
(seen as a Lie groupoid in the same way as $E$).

\begin{Lemma}
The restriction of the Fourier transform ${\mathcal F}_{\mu}$ to
$S(E)$ is a bijection onto $S(E^*)$, with inverse
\begin{equation}
({\mathcal F}_{\mu}^{-1}g)(\xi _x)=(2\pi)^{-p}\dint g(\theta_x)e^{i\langle 
\theta_x,\xi _x\rangle}d\mu_{*}^x(\theta_x). \ll{FTinv}
\end{equation}
\end{Lemma}
\begin{proof}
Compute in local coordinates.
\end{proof}

Using these two  lemmas we have
\begin{Proposition}
${\mathcal F}_{\mu}$ is an algebra isomorphism between $S(E)$ with the
convolution $\mbox{}^*$-algebra structure inherited from $C^*(E,\mu )$
and $S(E^*)$ with the $\mbox{}^*$-algebra structure borrowed from
$C_0(E^*)$.
\end{Proposition}

In the following lemma, which is used in the proof of Proposition
\ref{crochetEprop}, we list a series of properties of the Fourier
transform on $S(E)$.

\begin{Lemma}\label{lemafourier}
If $f,g\in S(E)$, $\phi \in S(E^*)$ and $a\in C^{\infty}(M)$, then 
\begin{enumerate}
\item $\left[ (a\circ \pi)f\right] ^{\hat{}}=(a\circ \pi )\hat{f}$;
\item $\dfrac{\d \hat{f}}{\d q_j}(\theta_x)= \widehat{\dfrac{\d f}{\d q_j}}
(\theta_x)+\dfrac{\d ln\mu _e}{\d q_j}(x)\hat{f}(\theta_x)$;
\item $\dfrac{\d \hat{f}}{\d \epsilon _i}(\theta_x)=-i(\xi _if)^{\hat{}}
(\theta_x)$;
\item $\dfrac{\d \check{\phi}}{\d \lambda _i}(\xi _x)= i(w _i\phi 
)^{\check{}}(\xi _x)$;
\item $i\theta_k\hat{f}(\theta_x)=\left( \dfrac{\d f}{\d \lambda _k}
\right) ^{\widehat{}}(\theta_x)$.
\end{enumerate}

\noindent Here $(\cdot)\hat{}$ is the Fourier transform ${\mathcal F}_{\mu}$ 
of $(\cdot)$ and $(\cdot)\check{}$ is the inverse Fourier transform  
${\mathcal F}^{-1}_{\mu}$ of $(\cdot)$.
\end{Lemma}

 The Schwartz functions do not form the only class of interest to us.
\begin{Definition}\ll{defpw}
The Paley--Wiener functions on $E^*$, denoted by $\CPW(E^*)$, consist all 
functions in $S(E^*)$ whose (inverse) Fourier transform is in $\cci(E)$.
\end{Definition}

We use this definition for $E=A(G)$; the Poisson algebra $A^0$ in
Theorems \ref{main1} and \ref{main2} is $\CPW(A^*(G))$. Writing the
Poisson bracket and the pointwise product in terms of the Fourier
transform, one quickly establishes that $A^0$ is indeed a Poisson
algebra (cf.\ section \ref{lsL}).
\section{Weyl quantization}\label{Weyl}
Having constructed the continuous field of \ca s called for in Theorem
\ref{main1} in section \ref{tg}, it remains to define a Poisson
algebra $A^0\subset C_0(A^*(G))$ and a map $\CQ:A^0\raw C^*(\hat{G})$,
or equivalently, a family of involutive maps $\q:A^0\raw C^*(G)$
satisfying (\ref{direq}).  These maps will, in addition, provide the
data of Theorem \ref{main2}.  We do so by a generalization of Weyl
quantization on $T^*\R^n$; cf.\ \ci{La3}.

 We pick a positive 1-density $\mu\in\cin(G_0,|\Om|^1(A(G)))$, with
 associated left Haar system $(\lm^x)_{x\in G_0}$ on $G$ (see section
 \ref{CLg}), left Haar system $(\hat{\lm}^{(\hbar,x)})_{(\hbar,x)\in
 \hat{G}_0}$ on $\hat{G}$ (see section \ref{tg}), and left Haar system
 $(\mu^x)_{x\in G_0}$ on $A(G)$ (see section \ref{Ft}).  We shall work
 with the concrete \ca s $C^*(G,\lm)$ and $C^*(\hat{G},\hat{\lm})$
 rather than with their intrinsically defined versions $C^*(G)$ and
 $C^*(\hat{G})$; see section \ref{CLg}. Moreover, since the argument
 below is the same for the reduced \ca s, we will not consider that
 case explicitly.

 Now choose some function
$\kp\in\cin(A(G),\R)$ with support in $V$ (cf.\ Proposition
\ref{tntoid}), equaling unity in some smaller tubular neighbourhood
of $G_0$, as well as satisfying $\kp(-\xi)=\kp(\xi)$ for all $\xi\in A(G)$.
\begin{Definition}\ll{Weylultimate}
Let $G$ be a Lie groupoid with Lie algebroid $A(G)$. We put
\begin{equation}
A^0=\CPW(A^*(G))\subset A_0=C_0(A^*(G)).
\end{equation}
 For $\hbar\neq
 0$, the  Weyl quantization of $f\in A^0$ is the element 
$\qw(f)\in\cci(G_1)$, regarded as a
 dense subalgebra of $C^*(G,\lm)$, defined by 
\bea
 \qw(f)(\Exp^W(\xi)) & = & |\hbar|^{-p}\kp(\xi)\F\inv_{\mu}f(\xi/\hbar)
\:\: \forall \xi\in V; \nn\\ 
\qw(f)(\gm) & = & 0 \:\: \forall \gm\notin W .
 \ll{qwfEW}
 \eea 
Here the Weyl exponential
 $\Exp^W:A(G)\raw G$ is defined in (\ref{ExpW}).
\end{Definition}

 This definition is possible by virtue of Proposition \ref{tntoid}.
(An analogous definition using $\Exp^L$ would not satisfy (\ref{real});
such a definition would generalize the Kohn--Nirenberg calculus of
pseudodifferential operators rather than the Weyl calculus.)
 By our choice of $A^0$, the operator $\qw(f)$ is
 independent of $\kp$ for small enough $\hbar$ (depending on $f$).
\begin{Proposition}\ll{T2}
  For each $f\in A^0$, the operator $\qw(f)$ of Definition
  \ref{Weylultimate} satisfies $\qw(f)^*=\qw(f^*)$, and the family
  $\{\qw(f)\}_{\hbar\in \R}$, with $\CQ^W_0(f)=f$, is a continuous
  cross-section of both continuous fields of \ca s in Theorem
  \ref{323}.
\end{Proposition}
\begin{proof}
It is immediate from (\ref{oldinv}) and (\ref{ExpW}) that for
real-valued $f\in A^0$ the operator $\qw(f)$ is self-adjoint in
$C^*(G,\lm)$; this implies the first claim.

Take $f\in\CPW(A^*(G))$. The function $\CQ^W(f)$ on
$\hat{G}_1$ that is defined by 
\bea
\CQ^W(f)(0,\xi) & = & \F_{\mu}\inv f (\xi); \nn \\
\CQ^W(f)(\hbar,\Exp^W(\xi)) & = &  \kp(\xi)\F\inv_{\mu}f(\xi/\hbar)
\:\: \forall \xi\in V; \nn\\ 
\CQ^W(f)(\gm) & = & 0 \:\: \forall \gm\notin W 
 \ll{qwfEd}
 \eea 
 and
 is in $\cci(\hat{G}_1)$; cf.\ Definition \ref{defnormoid}. 
In other words, $\CQ^W(f)$ is an element of $C^*(\hat{G},\hat{\lm})$.  

For $\hat{f}\in\cci(\hat{G})$, define the  restriction maps 
\bea \hat{\phv}_0(\hat{f}) : \theta & \mapsto & \F_{\mu}\hat{f}(0,\theta);
\nn \\ \hat{\phv}_{\hbar}(\hat{f}) : 
\gm & \mapsto & \hat{f}(\hbar,\gm) \:\: (\hbar\neq 0). \eea 
Then $\hat{\phv}_0$ is a surjective $\mbox{}^*$-homomorphisms from
$\cci(\hat{G}_1,\hat{\lm})$ to $\CPW(A^*(G),\mu)$, and each
$\hat{\phv}_{\hbar}$ is a surjective $\mbox{}^*$-homomorphisms from
$\cci(\hat{G}_1,\hat{\lm})$ to $\cci(G,|\hbar|^{-p}\lm)$. These maps
are contractive, and extend by continuity to surjective
$\mbox{}^*$-homomorphisms from $C^*(\hat{G},\hat{\lm})$ to
$C_0(A^*(G))$ and $C^*(G,|\hbar|^{-p}\lm)$, respectively.  However, in
Theorem \ref{main1} we have $A_0=C_0(A^*(G))$ and
$A_{\hbar}=C^*(G,\lm)$ for $\hbar\neq 0$. Hence the maps
$\phv_{\hbar}$ of Definition \ref{defcfca} should be taken as
\bea \phv_0(\hat{f}) : \theta & \mapsto & \F_{\mu}\hat{f}(0,\theta);
\nn \\ \phv_{\hbar}(\hat{f}) : 
\gm & \mapsto & |\hbar|^{-p}\hat{f}(\hbar,\gm)\:\: (\hbar\neq 0). 
\ll{phvhb} \eea 
These maps extend to surjective $\mbox{}^*$-homomorphisms from
$C^*(\hat{G},\hat{\lm})$ to $C_0(A^*(G))$ and $C^*(G,\lm)$,
respectively.  It follows that $A_{\hbar}\simeq
C^*(\hat{G},\hat{\lm})/\ker \phv_{\hbar}$ for all $\hbar\in\R$, and
that $\qw(f)=\phv_{\hbar}(\CQ^W(f))$. The proposition follows.
\end{proof}

We have now proved Theorem \ref{main1} up to Dirac's condition, and
turn to Theorem \ref{main2}.  We  in addition need to choose the
tubular neighbourhood $W$ of $G_0\subset G_1$ so that $\gm\gm'\subset W$
whenever $(\gm,\gm')\in G_2\cap (W\times W)$.

We define a semi-strict deformation quantization of $A^*(G)$ in the
following way. For given $f\in A^0=\CPW(A^*(G))$, we choose a cutoff
function $\ch_f$ on $\R$ that is 1 in a neighbourhood of 0 and has
support well inside the set of values of $\hbar$ for which $\CQ^W(f)$
as defined in (\ref{qwfEd}) is independent of $\kp$. It is clear from
the choice of $A^0$ that this can be done.  For each $\hbar\in\R^*$,
we then have:
\begin{itemize}
\item an involution $f^{*_{\hbar}}=\overline{f}$ (independent of $\hbar$);
\item a semi-norm $\| f\| _{\hbar}= \ch_f(\hbar)\| \qw(f)\|_{C^*(G,\lm)}$;
\item a product $\times_{\hbar}$ defined by the condition
$$\qw(f\times_{\hbar} g)=\ch_f\ch_g (\hbar) \qw(f) * \qw(g),$$ where
$*$ is the product in $C^*(G,\lm)$.
\end{itemize}

The existence of the product follows from a detailed but
straightforward analysis of the support properties of $\qw(f) *
\qw(g)$, leading to the conclusion that, seen as a function on $G_1$
for fixed $\hbar$, under the stated assumptions its support is
contained inside $W$; see \ci{Ra1}. Hence it can be pulled back using
$\Exp^W$ and can subsequently be Fourier-transformed so as to yield
the desired function $f\times_{\hbar} g$.

The fact that $*_{\hbar}$ is indeed an involution is a consequence of
the self-adjointness property (\ref{real}) of $\qw$. The same comment
applies to the $C^*$-property of the seminorm.  In general, these are
not norms, since for a given $f\neq 0$ there may well exist values of
$\hbar$ such that $\ch(\hbar) \qw(f)=0$, seen as a function on
$G(\hbar)\subset \hat{G}$.

The conditions defining a  semi-strict deformation quantization, except
(\ref{diracram}),  are now trivially satisfied
as a consequence of Theorem \ref{main1} as proved so far. A direct proof
of condition (b) in Definition \ref{semi}
is also immediate, based on 
 Theorem \ref{323}.
\section{The local structure of the Poisson bracket}\label{lsL}
In this section, taken from \ci{Ra1}, we express the structure of Lie
algebroids $E$, as well as the Poisson bracket on $\cin(E^*)$, in
local coordinates. This is of interest in itself, but in the context
of quantization it is a key tool for proving Dirac's condition
(\ref{direq}) or (\ref{diracram}).

Recall Definition \ref{defloid} of a Lie algebroid.  Let
$\left\{e_1,e_2,\ldots,e_p\right\}$ be a local frame on $U\subset M$
for $E$, and let $(q_1,\ldots,q_n,\lambda _1,\ldots,\lambda _p)$ be
local coordinates of $E$, with the $q_i$'s local coordinates for $M$
and the $\lambda_j$'s local coordinates for the fibers associated to
the frame $\left\{ e_1,e_2,\ldots,e_p\right\}$.  Then, locally, the
fact that $E$ is a Lie algebroid implies the existence of structure
functions $c_{ijk}$ and $a_{ij}$ in $C^{\infty}(U)$ such that
\bea
\left[ e_i,e_j\right] & = & \stackunder{k}{\dsum}c_{ijk}e_k; \\ 
\rho (e_i) & = & \stackunder{j}{\dsum}a_{ij}\dfrac{\partial}{\partial q_j}.
\eea

The Poisson bracket (\ref{pblieoid1}) - (\ref{pblieoid3}) of the
functions $\phi ,\psi \in C^{\infty}(E^*)$ is then locally given by
$$\left\{ \phi ,\psi \right\} =\stackunder{i,j}{\dsum}\rho
(e_i)(q_j)\left(
\dfrac{\d \phi }{\d \epsilon _i}\dfrac{\d \psi}{\d q_j}-\dfrac{\d \phi}{\d q_j}
\dfrac{\d \psi}{\d \epsilon _i}\right) +\stackunder{i,j}{\dsum} \widetilde{
\left[ e_i,e_j\right]} \dfrac{\d \phi}{\d \epsilon _i}\dfrac{\d \psi}{\d 
\epsilon _j};$$
recall the notation explained below (\ref{pblieoid3} ). 
Taking the value at the point $(x,\alpha )\in E^*_x$ and using the
structure functions of the Lie algebroid, we obtain
\begin{eqnarray}\label{crochetE*}
\left\{ \phi ,\psi \right\} (x,\alpha ) & = & 
\stackunder{i,j}{\dsum}a_{ij}(x)\left( \dfrac{\d \phi }{\d \epsilon _i}(x,
\alpha )\dfrac{\d \psi}{\d q_j}(x,\alpha )- \dfrac{\d \phi}{\d q_j}(x,\alpha )
\dfrac{\d \psi}{\d \epsilon _i}(x,\alpha )\right)  \nonumber \\
& + & \stackunder{i,j,k}{\dsum}c_{ijk}(x)\epsilon _k(\alpha ) \dfrac{\d \phi}{
\d \epsilon _i} (x,\alpha ) \dfrac{\d \psi}{\d \epsilon _j}(x,\alpha ).
\end{eqnarray}

Fix a positive 1-density $\mu \in C^{\infty }(M,|\Om |^1 (E))$. We
define a Poisson bracket on the dense subalgebra $S(E)$ of $C^*(E,\mu
)$, by transporting the Poisson bracket of $C^{\infty}(E^*)$ using the
Fourier transform ${\mathcal F}_{\mu}$. That is,
\begin{equation}\label{crochetE}
\left\{ f,g\right\} _{\mu} = {\mathcal F}^{-1}_{\mu} \left( \left\{ {\mathcal 
F}_{\mu}f,{\mathcal F}_{\mu}g\right\} \right).
\end{equation}
\begin{Proposition}\label{crochetEprop}
Eq.\ (\ref{crochetE}) defines a Poisson bracket on the convolution
algebra $S(E)$. The Poisson bracket of $f,g\in S(E)$ is explicitly
given by
\begin{eqnarray}\label{crosetE}
\left\{ f,g\right\} _{\mu}(\xi _x) & = & -i\stackunder{i,j}{\dsum}
a_{ij}(x) \left( \xi _if*\dfrac{\d g}{\d q_j} -\xi _ig*\dfrac{\d f}{\d q_j}
\right) (\xi _x) \nonumber \\
 & -  &   i\stackunder{i,j}{\dsum}a_{ij}(x)\dfrac{\d \ln \mu _e}{\d q_j}(x)
\left( \xi _if*g-\xi _ig*f\right) (\xi _x) \nonumber \\
 & +  & i\stackunder{i,j,k}{\dsum}c_{ijk}(x)\dfrac{\d }{\d \lambda _k}
\left( \xi _if*\xi _jg\right) (\xi _x).
\end{eqnarray}
\end{Proposition}

\begin{proof}
Eq.\ (\ref{crochetE*}) proves that $S(E^*)$ is stable under the Poisson 
bracket $\left\{ \cdot ,\cdot \right\}$, hence (\ref{crochetE}) 
is well defined. Since the Fourier transform is an algebra isomorphism between 
$S(E)$ and $S(E^*)$, it can  easily be shown that we have a Poisson 
algebra structure on $S(E)$. Eq.\ (\ref{crosetE}) is a consequence 
of Lemma \ref{lemafourier} and (\ref{crochetE*}).
\end{proof}

We now specialize to the case of relevance to us, where $E=A(G)$ is the
Lie algebroid of a Lie groupoid $G$. We   analyze 
 the local structure of $G$ 
in the neighborhood of a fixed point $x_0\in G_0\subset G$ in a 
convenient parametrization; also cf.\ \cite{nistor}.

The map $r$ is a submersion at $x_0\in G_0\subset G_1$, hence there exists
 an open neighborhood $U$ of 0 in $\R^n$, 
 an open neighborhood $V$ of 0 $\R^m$, and  parametrizations 
$\psi :U\times V\rightarrow G_1$ and $\varphi :U\rightarrow G_0$ such that
\begin{equation}\label{psibon}
\begin{array}{l}
1.\ \psi (0,0)=x_0;\\
2.\ r(\psi (u,v))=\varphi (u);\\
3.\ \psi (U\times \lbrace 0\rbrace )=\psi (U\times V)\bigcap G_0.
\end{array}
\end{equation}
 It follows from the first two conditions that $\varphi (u)=\psi
 (u,0)$. To the parametrization $\psi$ of the Lie groupoid $G$ one can
 associate a parametrization $\theta :U\times \R^m\rightarrow A(G)$ of
 the neighborhood $A(G)_{\varphi (U)}$ of the fiber $A(G)_{x_0}$ in
 $A(G)$, given by $\theta (u,v)=(\varphi (u),\dfrac{\partial
 \psi}{\partial v}(u,0)v)$.

For every $x_0\in G_0$ there exists a neighborhood $\psi (U\times
V)$ of $x_0$ in $G$ that is diffeomorphic to the neighborhood
$\theta (U\times V)$ of $(x_0,0)$ in $A(G)$ by $\alpha =\psi \circ
\theta ^{-1}$. Moreover, $\alpha (A(G) _x)\subset G^x$ for each $x\in
\varphi (U)$. This result can be formulated in a stronger form,
based on the existence of an exponential map for a Lie groupoid; see
section \ref{LgLa}.  Namely, taking $\alpha$ equal to the restriction
of $\Exp^L$ (cf.\ Definition \ref{defExpL}) to $V$, and $\alpha _x$ as
the restriction of $\alpha$ to $A(G) _x\bigcap V$, one achieves that
$\alpha (A(G)_x \bigcap V)=G^x\bigcap W$ and that $\alpha^{\prime}_x
(0)$ is the identity of $A(G) _x$.

The submersion $\sigma =\varphi ^{-1}\circ s\circ \psi:U\times V\rightarrow U$ 
is the local expression of the source map $s$ in the parametrization $\psi$.
 One has  $\sigma (u,0)=u$, since $\varphi (\sigma
(u,0))= s(\psi (u,0))=\psi (u,0)=\varphi (u)$. 
Similar expression may be given for the
multiplication and the inversion.

The following theorem gives a version of the Baker-Campbell-Hausdorff 
formula for  Lie groupoids. This  will enter the proof of Dirac's condition.
\begin{Theorem}\label{BCH}
Let $G$ be a Lie groupoid. Then 
\begin{enumerate}
\item $(\psi (u,v),\psi (u_1, w))\in G_2$ if and only if $u_1=\sigma (u,v)$. 
In that case, their product is of the form $\psi (u,v)\psi (\sigma (u,
v),w)=\psi (u,p(u,v,w))$, where $p:U\times V\times V\rightarrow V$ is a smooth
map which has the expansion 
\begin{equation}\label{bch}
p(u,v,w)=v+w+B(u,v,w)+O_3(u,v,w),
\end{equation} 
with $B$ bilinear in $(v,w)$, and $O_3(u,v,w)$ of the order of $\| (v,w)\|^3$.
\item  Let $(u,v)\in U\times V$ be such that $\psi (u,v)^{-1}\in \psi (U\times 
V)$. Then $\psi (u,v)^{-1}=\psi (\sigma (u,v),w)$, where $w$ satisfies 
$p(u,v,w)=0$. Moreover, $w=-v+B(u,v,v)+O_3(u,v)$, with $O_3(u,v)$ of the order
of $\| v\| ^3$.
\end{enumerate}
\end{Theorem}

\begin{proof}
We merely sketch the proof. For details cf.\ \ci{Ra4}.

1. Set $g=\psi (u,v)$ and $h=\psi (u_1,w)$. With $s(g)=\varphi (\sigma 
(u,v))$ and $r(h)=\varphi (u_1)$, we have $(g,h)\in G_2$ if and only if 
$u_1=\sigma (u,v)$. Also, $r(gh)=\varphi (u)$ implies the existence of a
unique $p(u,v,w)\in V$ such that $\psi (u,v)\psi (\sigma (u,v),w)=\psi 
(u,p(u,v,w))$. Hence we obtain a map $p:U\times V\times V
\rightarrow V$ that satisfies $p(u,0,w)=w$ and $p(u,v,0)=v$. 
Using these equations in a Taylor expansion of $p$ yields (\ref{bch}).

2. Similar to 1. 
\end{proof}

We can now give explicit formulae for the structure functions of the Lie
algebroid $A(G)$. First, remark that the family $\lbrace
e_1,e_2,\ldots,e_m
\rbrace$ defined by $e_i(\varphi (u))=\theta (u,f_i)$, $i=\overline{1,m}$, 
where $\lbrace f_1,f_2,\ldots,f_m\rbrace $ is the 
canonical basis of $\R^m$, is a 
frame of $A(G)$ on $\varphi (U)$. Also, recall that the anchor of 
$A(G)$ is given by $\rho =Ts$ (cf.\ Definition \ref{deflbroid}.2), 
that the local coordinate functions of
$G_0$ are $q_j=pr_j\circ \varphi ^{-1}$, and that $B_1,\ldots,B_m$ are
the coordinates of $B:U\times
V\times V\rightarrow V$ in the 
 base $\left\{ f_1,f_2,\ldots,f_m\right\}$ of $\R ^m$.
Finally, recall that $a_{ij}=\rho (e_i)(q_j)$, and that
 the $c_{ijk}$ are given by
$\left[ e_i,e_j\right] =\dsum c_{ijk}e_k$.

\begin{Proposition}\label{329}
For each $u\in U$, the structure functions of the Lie algebroid $A(G)$ are
given by $a_{ij}(\varphi (u))=\dfrac{\partial \sigma _j}{\partial v_i}(u,0)$
and $c_{ijk}(\varphi (u))=B_k(u,f_i,f_j)-B_k(u,f_j,f_i)$.
\end{Proposition}

\begin{proof}
Direct calculations. See \cite{Ra1}, \cite{Ra4} for details.
\end{proof}

We now return to the Poisson bracket on $A(G)$.  Recall Proposition
\ref{crochetEprop}, and put $E=A(G)$. The next proposition gives a
local expression of the Poisson bracket on $S(A(G))$.

Let $\lambda \in C^{\infty}(G_0,|\Om |^1(A(G)))$ be a positive
1-density and $\Lambda$ the local expression in the parametrization $\theta$ 
of the function associated to $\lambda$ in the local frame 
$\left\{ e_1,e_2\ldots,e_m\right\}$. 
\begin{Proposition}\label{croset G}
For $f,g\in S(A(G))$, let $h=\left\{ f,g\right\}_{\lambda}$ be the
Poisson bracket of $f$ and $g$ given by (\ref{crosetE}).  Denote the
local expressions of $f,g,h$ in the parametrization $\theta$ by
$F,G,H$, respectively. Then \\ $H(u,v)= i\dint F(u,w)G(u,v-w)
\stackunder{j}{\dsum}\left[ B_j(u,w,f_j)-B_j(u,f_j,w)\right] \Lambda (u)dw$
\\ $+ i\dint \stackunder{k}{\dsum}\left[ 
B_k(u,w,v)-B_k(u,v,w)\right] F(u,w)\dfrac{\partial G}{\partial
v_k}(u,v-w)\Lambda (u)dw$ \\ $- i\stackunder{i,j}{\dsum}
\dfrac{\partial
\sigma _j}{\partial v_i}(u,0)
 \dfrac{\partial \Lambda}{\partial u_j}(u)\dint w_i 
\left[ F(u,w)G(u,v-w)-G(u,w)F(u,v-w)\right]dw$ \\
$- i\stackunder{i,j}{\dsum} \dfrac{\partial \sigma _j}{\partial
v_i}(u,0)
\dint w_i \left[ F(u,w) \dfrac{\partial G}{\partial u_j}(u,v-w)-
G(u,w)\dfrac{\partial F}{\partial
u_j} (u,v-w)\right] \Lambda (u)dw.$
\end{Proposition}

\begin{proof}
Replacing the structure functions in (\ref{crosetE}) by their
expressions calculated in Proposition \ref{329}, we can directly
identify the first two lines of (\ref{crosetE}) with the last two in
the formula of $H(u,v)$. Denote the expression in the last line of
(\ref{crosetE}) by $l$. We then have
\begin{eqnarray} 
l(\theta (u,v)) & = &
 -\stackunder{i,j,k}{\sum}c_{ijk} (\varphi (u))\dfrac{\d}{\d 
v_k} \left( \dint w_iF(u,w)(v_j-w_j)G(u,v-w)\Lambda (u)dw\right) \nn \\
& = &
\stackunder{i,j}{\sum}\dint (
B_j(u,f_j,w_if_i)-B_j(u,w_if_i,f_j)) F(u,w)G(u,v-w)\Lambda (u)dw \nn \\
 & + &
\stackunder{i,j,k}{\sum}\dint \lbrack 
B_k(u,(v_j-w_j)f_j,w_if_i)-B_k(u,w_if_i,(v_j-w_j)f_j)\rbrack \nn \\
& \cdot & F(u,w)
\dfrac{\partial G}{\partial v_k}(u,{v-w})\Lambda (u)dw \nn \\
 & = &
\stackunder{j}{\sum} \int ( 
B_j(u,f_j,w)-B_j(u,w,f_j)) F(u,w)G(u,v-w)\Lambda (u)dw \nn \\
 & + &
\stackunder{k}{\sum}\dint ( B_k(u,v-w,w)-B_k(u,w,v-w)
) F(u,w)\dfrac{\partial G}{\partial v_k}(u,v-w)\Lambda (u)dw. \nn
\end{eqnarray}

By bilinearity one has $B_k(u,w,v-w)-B_k(u,v-w,w)=B_k(u,w,v)-
B_k(u,v,w)$; this finishes the proof.
\end{proof}
\section{Proof of Dirac's condition}\label{ssc}
We now use the results of the preceding section to prove (\ref{direq}) and
(\ref{diracram}) in the case at hand. We start with two lemmas.

 Let $\psi$ be a parametrization of $G$ as in section \ref{lsL}, and denote 
the function associated to $\lambda$
in the local frame of $TG$ generated by $\psi$
by $\lambda _0$. Put $\Lambda = \lambda _0\circ \psi$.
\begin{Lemma}\label{densequiva}
One has $\lambda _0(\gm )=\dfrac{\lambda _0(s(\gm ))}{|det J_{L_{
\gm }}(s(\gm ))|}$ for every $\gm$ in the image of the 
parametrization $\psi$.
\end{Lemma}
\begin{proof}
The left invariance of $\lambda $ means  $\lambda (\gm )=
\gm \cdot \lambda (s(\gm ))$. It only remains  to write the action of $G$ on 
$|\Om |^1(\ker Tr)$ in the local parametrization $\psi$. We leave this to the
reader.
\end{proof}
\begin{Lemma}\label{derivdens}
Let $\mu _{u,v}(t)=\lambda _0(\psi (u,tv))$. Then 

\centerline{$\mu '_{u,v}(0)=\stackunder{i,j}{\sum }\dfrac{\d \Lambda}{\d u_i}
(u,0)\dfrac{\d \sigma _i}{\d v_j}(u,0)v_j-\lbrack B_1(u,v,f_1)+
+\cdots+B_m(u,v,f_m)\rbrack \Lambda (u,0).$}
\end{Lemma}

\begin{proof}
By Lemma \ref{densequiva},  
$\mu _{u,v}(t)=\dfrac{\lambda _0(\psi 
(\sigma (u,tv),0))}{|det J_{L_{\psi (u,tv)}}
(\psi (\sigma (u,tv),0))|}$.
Write 
\bea
a_{u,v}(t)& = & \lambda _0(\psi (\sigma (u,tv),0));\nn \\
b_{u,v}(t) & = & |det J_{L_{\psi (u,tv)}}(\psi (\sigma (u,tv),0))|.\nn
\eea 
We have $a_{u,v}=\Lambda (\sigma (u,tv),0)$ and 
$a'_{u,v}(0)=\stackunder{i,j}{\sum }
\dfrac{\d \Lambda}{\d u_i}(u,0)\dfrac{\d \sigma _i}{\d v_j}(u,0)v_j$.

The local expression of the multiplication map $L_{\psi (u,tv)}:
G^{\varphi (\sigma (u,tv))}\rightarrow G^{\varphi (u)}$ is given by
$p(u,tv,\cdot )$ and this shows, after some calculation, 
$$b_{u,v}(t)=\left| 
\begin{array}{cccc}
1+tB_1(u,v,f_1) & tB_1(u,v,f_2) & \cdots & tB_1(u,v,f_m)\\
tB_2(u,v,f_1) & 1+tB_2(u,v,f_2) & \cdots & tB_2(u,v,f_m)\\
\cdots & \cdots & \cdots& \cdots\\
tB_m(u,v,f_1) & tB_m(u,v,f_2) & \cdots & 1+tB_m(u,v,f_m)
\end{array}
\right|. $$

Taking the derivative we obtain $b'_{u,v}(0)=B_1(u,v,f_1)+B_2(u,v,f_2)+
\cdots+B_m(u,v,f_m)$, and to finish the proof remark that $a_{u,v}(0)=\Lambda
(u,0)$, $b_{u,v}(0)=1$. 
\end{proof}

Now let $(\gm ,\hbar)\in\hat{G}$, with $\hbar\not= 0$.  The smooth 
structure in a neighborhood of $(\gm ,\hbar)$ is given by a parametrization 
of the form $\hat{\psi}=\psi \times id:U\times \R \rightarrow \hat{G}$, 
where $\psi :U\rightarrow G$ is a parametrization of $G$ in a neighborhood of
$\gm$ and $U$ is an open subset of $\R^{n+m}$. Write $\alpha = \hat{\psi}^{
-1}:\hat{\psi }(U\times \R)\rightarrow U\times \R$, and define an element of
 $T_{(\gm ,\hbar)}\hat{G}$ by 
$$\dfrac{\partial}{\partial \hbar}\hat{f}|_{(\gm ,\hbar)}=
\dfrac{\partial \hat{f}}{\partial \alpha _{n+m+1}}(\gm ,\hbar).$$ 
Here $\hat{f}\in C^{\infty}(
\hat{G})$. This definition is independent of the 
parametrization $\psi$ of $G$.  Thus we obtain a vector field
$\dfrac{\partial}{\partial \hbar}\in
C^{\infty}(\hat{G},T\hat{G})$.

The following result is the key lemma towards the proof of Dirac's condition.
 \begin{Lemma}\label{fonctions}
Let $\hat{f},\hat{g}\in C_c^{\infty}(\hat{G},
\hat{\lambda})$ (seen as a convolution algebra), 
 and write $\hat{f}_0=\hat{f}_{|_{\hat{G}(0)}}$,
 $\hat{g}_0=\hat{g}_{|_{\hat{G}(0)}}$ for their the restrictions to
 the Lie algebroid $A(G)$. Then $$\dfrac{\partial}{\partial
 \hbar}\lbrack \hat{f},\hat{g}\rbrack _{|_{\hbar=0}}= i
\lbrace \hat{f}_0,\hat{g}_0\rbrace _{\mu}.$$
\end{Lemma}

\begin{proof}
The proof is only sketched; for details cf.\ \cite{Ra1,Ra3}.  Using a
partition of unity, one may assume that $\hat{f}$ and $\hat{g}$ have
their support contained in the image of the parametrization
$\hat{\psi}$. Let $\hat{k}$ be the convolution of $\hat{f}$ and
$\hat{g}$, and let $K$ be the expression of $\hat{k}$ in the
parametrization $\hat{\psi}$.  After some calculation and the use of
(\ref{bch}) we obtain
\bea
K(u,v,\hbar)& = & \dint F(u,w,\hbar)G(\sigma (u,\hbar w),v-w+\hbar
B(u,w,w)\nn \\ & - & \hbar B(\sigma (u,\hbar w),w,v)+O(\hbar ^2),\hbar
)\mu _{u,w}(\hbar )dw. \nn
\eea
This leads to an expression for the commutator
$\hat{f}*\hat{g}-\hat{g}*\hat{f}$. We then differentiate $K$ with
respect to $\hbar $ at $\hbar =0$, use Lemma \ref{derivdens} and some
changes of variables, and eventually recover the local expression of
the Poisson bracket as given in Proposition \ref{croset G}.
\end{proof}

We are now in a position to prove Dirac's condition (\ref{diracram})
and (\ref{direq}).  We only deal with the former; the latter is proved
in exactly the same way. The essence of the proof is that Dirac's
condition follows from the continuity condition (b) in Definition
\ref{semi}; in the context of Definition \ref{defqua} this condition is
 satisfied as a 
consequence of the definition of a continuous field of \ca s.

Define $\hat{V}\subset\hat{A(G)}$ as $V$ in Proposition \ref{tntoid},
but now for the tangent groupoid $\hat{G}$ rather than $G$.  For
$f,g\in C_c^{\infty}(A(G) )$, let $\varphi :\hat{V}\rightarrow \C$ be
defined by
\bea
\varphi (x,X,0)& = & 0; \nn \\
\varphi (x,X,\hbar )& = & \dfrac{(f\times _{\hbar}g)(x,X)-
(g\times _{\hbar} f)(x,X)}{\hbar }-i \left\{ f,g\right\} (x,X) 
\forall\hbar \neq 0.\nn
\eea
  Lemma \ref{fonctions} shows that $\varphi \in C^{\infty}
(\hat{V})$, and arguments like those  in the discussion of the 
product $f\times _{\hbar}g$ in section \ref{Weyl} show
 that supp$\, \varphi$ is  compact in $\hat{V}$.
Write $\hat{\varphi}=\varphi \circ \hat{\alpha} ^{-1}
\in C_c^{\infty}(\hat{\alpha}(\hat{V}))$. One may regard 
$\varphi$ as an element of $C_c^{\infty}(A(G) \times \R )$ and
$\hat{\varphi}$ as an element of $C_c^{\infty}(\hat{G})$ by
extending these functions by 0 outside $\hat{V}$ and
$\hat{\alpha}(\hat{V})$, respectively.  Remark that for every
$\hbar$ one has function $\varphi (\cdot ,\hbar )\in A^0$.  Applying
Theorem \ref{323} and recalling the notation (\ref{Ghbar}), where we
look at the $G(\hbar)$ as subgroupoids of $\hat{G}$, one then has
$$\stackunder{\hbar \rightarrow 0}{\lim}\left\| \varphi (\cdot ,\hbar
)\right\| _{\hbar} =\stackunder{\hbar \rightarrow 0}{\lim}\left\|
\hat{\varphi}_{|_{G(\hbar )}}
\right\| _{C^*(G(\hbar ),\lm)}= \left\| \hat{\varphi}_{|_{G(0)}}
\right\| _{C^*(G(0),\mu)}=0.$$

This finishes the proof of Dirac's condition. Having giving the other
parts of the proof in section \ref{Weyl}, this finishes the proof of
Theorems \ref{main1} and \ref{main2}.
\section{Examples and comments}\label{Ec}
Since a vast number of interesting \ca s are defined by some Lie
groupoid, one obtains a large reservoir of examples of our theorems,
of which in this section we merely scratch the surface.
\begin{Example} \label{Lie}
Quantization of the Lie-Poisson structure on a dual Lie
algebra
\end{Example}

This example was already introduced in the Introduction. 
A Lie group is a Lie groupoid with $G_0=e$.  A left-invariant Haar
measure on $G$ provides a left Haar system; the ensuing convolution
algebra $C^*(G)$ is the usual group $C^*$-algebra. 
The Poisson structure on the dual Lie algebra $A^*(G)$ is 
 the well-known  Lie-Poisson structure \ci{MR,Vai}. 
No connection is needed to define the exponential map, and one has 
\begin{equation}
\Exp^L(X)=\Exp^W(X)=\Exp(X),\ll{expWG} \end{equation} 
where $X\in\g$ and $\Exp:\g\raw G$ is the usual exponential map.  We
obtain a semi-strict or strict deformation quantization of $A^*(G)$ for
any Lie group $G$, but the situation is particularly favourable when
$G$ is exponential (in that $\Exp$ is a diffeomorphism).  One may then
omit the cutoff functions $\kp$ and $\ch$ in section \ref{Weyl}, and
$\qw(f)$ in (\ref{qwfEW}) is simply given by
\begin{equation}
\qw(f):\Exp(X)\raw(2\pi\hbar)^{-n}\int_{\g^*} 
e^{i\la\theta,X \ra/\hbar}f(\theta)d^n\theta.\ll{defqhc}
\end{equation} This is precisely Rieffel's prescription \ci{Ri4}; his
assumption that $G$ be nilpotent is now seen to be unnecessary in
order to obtain a strict deformation quantization.

Moreover, in the exponential case our semi-strict quantization is
easily shown to be strict, since the semi-norms $\|\cdot\|_{\hbar}$
are now norms.  More generally, if a Lie groupoid is diffeomorphic to
its Lie algebroid then the semi-strict deformation quantization we
have given is always strict; see \ci{Ra1}.
\begin{Example}
Transformation group \ca s
\end{Example}
Let a Lie group $H$ act smoothly on a manifold $M$. The transformation
group\-oid (or action groupoid) $G=H\x M$ is defined by the operations
$s(x,m)=x\inv m$ and $r(x,m)=m$, so that 
$((x,m),(y,m'))\in G_2$ when $m'=x\inv m$. In that case, $(x,m)\cdot
(y,x\inv m)=(x y,m)$. The inclusion is $m\mapsto (e,m)$, and  the
inverse is $(x,m)\inv=(x\inv,x\inv m)$.

Each left-invariant Haar measure $dx$ on $H$ leads to a left Haar
system on $G$. The corresponding groupoid \ca\ is the usual
transformation group \ca\ $C^*(G)=C^*(H,M)$, cf.\ \ci{Re1}.

The Lie algebroid $\H\x M$ (where $\H$ is the Lie algebra of $H$) 
is a trivial bundle over $M$, with
 anchor $\rh(X,m)=-\xi_X(m)$ (the fundamental vector field on $M$
 defined by $X\in\H$). Identifying sections of $\H\x M$ with
 $\H$-valued functions $X(\cdot)$ on $M$, the Lie bracket on $\cin(M,\H\x
 M)$ is 
\begin{equation} [X,Y]_{\H\x M}(m)=[X(m),Y(m)]_{\H}+\xi_Y X(m)-\xi_X
 Y(m). \ll{LBactioloid} \end{equation} The associated Poisson bracket coincides
 with the semi-direct product bracket defined in \ci{KM}.

The trivial connection on $\H\x M\raw M$ yields the exponential maps
\bea \Exp^L(X,m) & = &
(\Exp(X),m) ;\ll{expLAG}\\ \Exp^W(X,m) & = & (\Exp(X),\Exp(\half
X)m).\ll{expWAG} \eea 
The cutoff function $\kp$ in (\ref{qwfEW}) is independent
of $m$, and coincides with the function appearing in Example
\ref{Lie}.  For small enough $\hbar$, a function $f\in \CPW(\H^*\x M)$
is then quantized by \begin{equation} \qw(f):(\Exp(X),m)
\raw(2\pi\hbar)^{-n} \int_{\H^*}
e^{i\la\theta,X\ra/\hbar}f(\theta,\Exp(-\half X)m)d^n\theta. \ll{qwongact} 
\end{equation}
When $H=\R^n$ and $M$ has a $H$-invariant measure, the map $f\raw
\qw(f)$ is equivalent to the deformation quantization considered
by Rieffel \ci{Ri2}, who already proved that it is strict.

Finally, when $H$ is exponential we are in the situation discussed at
the end of the previous example, so that our semi-strict deformation
quantization is strict.
\begin{Example}
Weyl quantization on $T^*\R^n$
\end{Example}
As already remarked in the Introduction, the tangent bundle $TM$ of a
manifold $M$ is the Lie algebroid of the pair groupoid $G=M\x
M$. Hence our formalism produces a (semi) strict deformation
quantization of the cotangent bundle $T^*M$ of any manifold with
linear connection.

In order to understand Weyl quantization in the light of our
formalism, we must change some signs: we add minus signs in front of
the right-hand sides of (\ref{pblieoid2}) and (\ref{pblieoid3}), and
change $f(\xi/\hbar)$ in (\ref{qwfEW}) to $f(-\xi/\hbar)$ (cf.\
\ci{La3}) for the rationale behind these signs). For $M=\R^n$ with
flat connection (\ref{qwfEW}) then simply becomes
\begin{equation}
\qw(f)\Ps(x)=(2\pi\hbar)^{-n} \int_{T^*\R^n} 
e^{ip(x-y)/\hbar}f(p,\half(x+y) )\Ps(y)d^npd^ny, \ll{defweylq} \end
{equation} where $\Ps\in L^2(\R^n)$; we here use the fact that
$C^*(\R^n\x\R^n)=\B_0(L^2(\R^n))$.  This is precisely Weyl's original
prescription, now written in a form that emphasizes its geometric
origin. For $\half(x+y)$ is the midpoint of the geodesic from $x$ to
$y$, and $x-y$ is a tangent vector to the geodesic at this midpoint.
Looked at in this way, Weyl quantization may easily be generalized to
manifolds with connection \ci{La1,La3}, and forms a special case of
what we have called Weyl quantization for general Lie groupoids in
this paper. This answers Rieffel's Question 20 in \ci{Ri5}.

 The fact that this quantization is strict, and in particular
 satisfies (\ref{direq}), had already been proved by Rieffel \ci{Ri6}.
 The associated continuous field of \ca s has fibers $A_0=C_0(\tsr)$
 and $A_{\hbar}=\B_0(L^2(\R^n))$ for $\hbar\neq 0$.  The \ca\ $C$ in
 Definition
\ref{defcfca} is $C^*(H_n)$, the  group algebra of
the simply connected Heisenberg group on $\R^n$; also see \ci{ENN1}. This is
indeed the  \ca\ of the tangent groupoid of $\R^n$.

When $M$ be Riemannian and exponential (in that the exponential map is
a diffeomorphism between the tangent bundle $TM$ and the pair groupoid
$M\times M$), our semi-strict deformation quantization is strict
\ci{Ra1}.  As is well known, for $M=\R^n$ with flat metric the product
$\times_{\hbar}$ is then equal to the Moyal product
\ci{Mo,BFFLS}. For a detailed analysis of this situation in the setting of the
present paper see \cite{CCFGRV}.

We close with two remarks.
\begin{remark}
There are clear connections between index theory (in the sense of
Atiyah--Singer), $C^*$-algebraic K-theory, and quantization; see,
e.g., \ci{Fedosov,Higson,ENN2,Rosenberg}. Moreover, Connes \ci{connes}
discovered a beautiful proof of the index theorem that is based on the
tangent groupoid of a manifold.  

Let $M$ be a compact Riemannian manifold, with associated cosphere bundle
$S^*M$ and \ca\ of bounded classical pseudodifferential operators
$\Ps^0(M)$.
The interpretation of index theory
in terms of the K-theory of \ca s comes from the short exact sequence
of \ca s
\begin{equation}
\ll{ses} 0\raw \B_0(L^2(M))\raw \Ps^0(M)\stackrel{\sg}{\raw} C(S^*M)\raw 0;
\end{equation}
here $\sg$ is the symbol map.
Any short exact sequence $0\raw J\raw A\raw B\raw 0$ leads to a
connecting map $\partial: K^1(B)\raw K_0(J)$, so that for (\ref{ses})
one obtains $$
\partial: K_1(C(S^*M))=K^1(S^*(M))\raw K_0(\B_0)={\Bbb Z}.
$$
The analytic index of an elliptic pseudodifferential operator $P\in 
\Ps^0(M)$ is given by $\partial([\sg(P)])\in {\Bbb Z}$, where, for
$f\in C(X)$,  $[f]$ is the class in $K^1(X)$ defined by $f$.
One may prove the Atiyah--Singer index theorem from
the existence of a strict deformation quantization $\q:C_0(T^*M)\raw
\B_0(L^2(M))$; cf.\ \cite{Higson}. 

Seen in the light of Lie groupoids and quantization, its is obvious
that the above data may be generalized \ci{MP} by replacing the pair groupoid
on $M$ by an arbitrary Lie groupoid $G$.Then $\B_0(L^2(M))$ is 
replaced by $C^*(G)$, the cosphere bundle $S^*M$ is replaced by
a cosphere bundle $S^*A(G)$ in $A^*(G)$, and
the appropriate generalization $\Ps^0(G)$ 
(with some abuse of notation) of
$\Ps^0(M)$ has recently been defined as well \ci{nistor,MP}.
Thus (\ref{ses}) is generalized to
\begin{equation}
0\raw C^*(G)\raw \Ps^0(M)\stackrel{\sg}{\raw} C_0(S^*A(G))\raw 0;
\end{equation}
see \cite{MP}.

On may therefore expect that our strict deformation quantization 
of $A^*(G)$ eventually leads to an index theorem, where the index
now takes values in $K_0(C^*(G))$.
\end{remark}
\begin{remark}
There is no analogue of Lie's Third Theorem for Lie algebroids.
Examples given in \cite{AM} show the existence of Lie algebroids that
are not associated to any Lie groupoid. For an arbitrary Lie algebroid
$A(G)$, Pradines proved (cf.\ \cite{P3}) that one can merely construct a
local Lie groupoid $G$ which has $A(G)$ as its Lie algebroid.  In this
paper, we always assume that we have a given Lie groupoid $G$ with Lie
algebroid $A(G)$.  In order to use the ideas from this paper to
quantize all Lie algebroids, we would need to extend two
constructions, the $C^*$-algebra and the tangent groupoid which can be
associated to every Lie groupoid, to the context of local Lie
groupoids. The construction of the tangent groupoid has been extended
to local Lie groupoids in the context of pseudo-differential operators
by Nistor, Weinstein and Xu (cf.\ \cite{nistor}). The missing
ingredient in order to extend our results to arbitrary Lie algebroids
is therefore the construction of the $C^*$-algebra of a local Lie
groupoid.
\end{remark}
\noindent {\it Acknowledgement.} B.R. is greatly 
indebted to Jean Renault for many helpful discussions.

\bibliographystyle{amsalpha}

\end{document}